\documentstyle[12pt]{article}

\newcommand{\DR}{\mbox{{\footnotesize{$\overline{{\rm DR}}$}}}}
\newcommand{\MS}{\mbox{{\footnotesize{$\overline{{\rm MS}}$}}}}
\def\gsim{\kern.4em\raise.3ex
\hbox{$>$\kern-.75em\lower1ex\hbox{$\sim$}}\kern.4em}
\def\lsim{\kern.4em\raise.3ex
\hbox{$<$\kern-.75em\lower1ex\hbox{$\sim$}}\kern.4em}
\def\psla{p\kern-.45em/}

\renewcommand{\thefootnote}{\fnsymbol{footnote}}

\begin{document}

\setcounter{page}{0}
\thispagestyle{empty}

\begin{flushright}
KEK--TH 520 \\
SLAC--PUB--7558 \\
TU--522 \\
June 1997 \\
\end{flushright}

\vspace{2cm}

\begin{center}
{\large\bf Slepton Production as}\\
\vskip 0.5cm
{\large\bf a Probe of the Squark Mass Scale}
\end{center}
\baselineskip=32pt

\centerline{Mihoko M. Nojiri,$^a$ Damien M. Pierce,$^b$ 
and Youichi Yamada$^c$}

\baselineskip=22pt

\begin{center}
\footnotesize\it
$^a$\,KEK Theory Group, Oho 1-1, Tsukuba, Ibaraki, 305 Japan \\

$^b$\,Stanford Linear Accelerator Center, Stanford University,
Stanford, California 94309, USA\\

$^c$\,Department of Physics, Tohoku University, Sendai, 980-77, Japan
\end{center}

\vspace{1cm}

\begin{abstract}
We investigate an important radiative correction to the slepton and
chargino production processes.  We consider a supersymmetric spectrum
with a large splitting between the squark and slepton masses.  In this
case, in the effective theory below the squark mass threshold, the
supersymmetric Slavnov-Taylor identities which enforce the equality of
the gauge and gaugino couplings are violated.  The gaugino propagators
receive potentially large corrections due to virtual quark/squark loop
effects.  We compute the full one-loop (s)quark corrected slepton
production cross-sections.  The $t$-channel one-loop scattering
amplitudes are factorized into an effective chargino/neutralino mass
matrix and an effective fermion-sfermion-gaugino coupling. The
difference between the effective gaugino coupling and the gauge
coupling is proportional to $\log (M_{\tilde{Q}}/m_{\tilde{\ell}})$ in
the large squark mass limit.  We find that the one-loop corrected
slepton production cross-sections can depend on the squark mass
strongly, up to $9\% \times
\log_{10}(M_{\tilde{Q}}/m_{\tilde{\ell}})$.  We investigate the squark
mass sensitivity of the slepton cross-section measurements at a future
linear collider.  For sneutrino production accessible at $\sqrt{s}=
500$ GeV there can be sensitivity to squark masses at or larger than 1
TeV.
\end{abstract}

\vspace{2cm}

\vfill

{\noindent\em D.M.P. is supported by Department of Energy contract
DE--AC03--76SF00515. M.M.N. is supported in part by Grant in aid 
for Science and Culture of Japan (07640428, 09246232),
and JSPS Japanese-German Cooperative Science Program.}

\pagebreak

\normalsize\baselineskip=15pt
\setcounter{footnote}{0}
\renewcommand{\thefootnote}{\arabic{footnote}}

\section{Introduction}

Supersymmetry is an attractive possibility beyond the standard model.
Because of the relations supersymmetry imposes among the dimensionless
couplings, the quadratic divergences in the Higgs sector are cut-off
by the superpartner mass scale.  The cancellations stabilize the
hierarchy between the Planck scale and the weak scale.  The minimal
supersymmetric standard model (MSSM) is consistent with gauge coupling
unification suggested by grand unified theories.  Also, it is
interesting that one of the hallmarks of supersymmetry, a light Higgs
boson ($m_h\lsim130$ GeV), is favored by global fits to precision
electroweak data \cite{HIGGS}.

If low energy supersymmetry exists, superpartners will be copiously
produced at the next generation of supercolliders. For example, the
Large Hadron Collider (LHC) could produce thousands of gluino pairs in
a single day. Each gluino would then, in a typical scenario, decay to
the lightest supersymmetric particle (LSP) through a cascade of
charginos, neutralinos, jets and leptons. Detailed information about
the superpartner masses can be determined from such processes
\cite{HPSSY}. A future Linear Collider (LC) would likewise prove to be
an invaluable tool in measuring the parameters of the
soft-supersymmetry breaking Lagrangian, as precision measurements of
the superpartner spectrum, cross-sections, and branching ratios are
possible \cite{JLC1,TSUKA,FPMT,NOJIRI,BMT,NFT}.  The constrained
kinematics, tunable center-of-mass energy, and beam polarization allow
for direct and model independent interpretations of the
measurements. Also, in many processes the backgrounds are quite small.

In this paper we examine the prospect of testing supersymmetry via a
precise measurement of the lepton-slepton-gaugino vertex. The linear
collider provides a suitably clean experimental environment.  Among
the relations which account for the cancellations of quadratic
divergences, supersymmetry relates the lepton-slepton-gaugino
coupling to the usual gauge coupling.

Although bare (or \DR) couplings enjoy the relations imposed by
supersymmetry, the effective gauge and gaugino couplings are not equal
because supersymmetry is broken. In particular, all non-singlet
nondegenerate supermultiplets such as the quark-squark supermultiplets
contribute to the splitting.  Even arbitrarily heavy split multiplets
will contribute, although the contribution will be suppressed by the
heavy mass scale.  Hence, measurements of the type we consider here
not only provide for detailed tests of supersymmetry, but can also
elucidate important features of the scale and pattern of supersymmetry
breaking \cite{CFPa,CFPb,RKS,PT}.

For example, the (s)quark contribution to the splitting of the U(1)
and SU(2) gaugino/gauge (s)lepton couplings grows logarithmically with
the squark mass, as
\begin{equation}
\label{ln msq}
{\delta {g_Y} \over {g_Y}} \simeq {11g_Y^2\over48\pi^2}\ln\left(
{M_{\tilde Q}\over m_{\tilde \ell}}\right)\ ,
\qquad\qquad {\delta g_2\over g_2} \simeq
{3g_2^2\over16\pi^2}\ln\left(M_{\tilde Q}\over m_{\tilde\ell}\right)\ .
\end{equation}
This correction is obtained by evolving the couplings according to the
renormalization group equations (RGE's) of the effective theory
\cite{C} below the squark mass threshold.  When $M_{\tilde
Q}/m_{\tilde\ell}\simeq 10$ the correction to the SU(2) (U(1))
coupling is about 2\% (0.7\%). This gives rise to an enhancement of
the $t$-channel slepton or gaugino production cross-section of about
8\% (2.8\%).  If large statistics are available and systematic errors
can be controlled, we can (assuming the MSSM) constrain the squark
mass scale by this measurement. Notice that a modest hierarchy between
the squark and slepton masses is typical of the simplest gauge
mediated models, which predict $M_{\tilde Q}/m_{\tilde \ell_R} \simeq
6$.

As another example, it has been proposed that the scalar masses of the
first two generations might be very heavy (e.g. 20 TeV) so
that the FCNC constraints on the scalar masses are alleviated
\cite{CKN}. This large hierarchy in the superpartner spectrum will be
reflected in the splitting of the gaugino/gauge couplings of the third
generation (s)fermions.

We restrict our attention to the measurement of the first generation
lepton-slepton-gaugino coupling at an $e^-e^+$ linear collider. Much
study has been undertaken to determine how accurately we can expect to
measure these couplings. Two relevant processes to consider are
chargino production and slepton production. These processes involve
the lepton-slepton-gaugino coupling by way of the $t$-channel exchange
of a sfermion or gaugino. The measurement of the gaugino coupling via
the chargino production process has been considered in
Ref. \cite{FPMT}.  In that study the authors worked under the
assumption that only chargino production was available.  It was shown
that the $\tilde W\tilde{\nu}_e e_L$ coupling could be determined to
about 20\%.  Incorporating knowledge about the sneutrino mass allows
for an improved ${\cal O}(2$-$3\%)$ determination of the $\tilde
W\tilde{\nu}_e e_L$ coupling \cite{CFPb}.  A more precise determination
is possible by studying slepton production. In Ref. \cite{NFT} it was
demonstrated that by measuring $\tilde e_R^-\tilde e_R^+$ production
the U(1) $\tilde B\tilde e e$ coupling could be determined to about
$1\%$. Note that this is the same order of magnitude as the splitting
due to heavy squarks (see Eq.~(\ref{ln msq})). In Ref.~\cite{CFPb} it
was estimated that a ${\cal O}(0.3\%)$ measurement of the U(1) gaugino
coupling could be achieved by measuring $\tilde e_R^-\tilde e_R^-$
production at an $e^-e^-$ collider.

We perform a full one-loop calculation of the slepton production
cross-section within the MSSM. Since we are interested in the
sensitivity to the squark mass scale, we assume the squark mass
spectrum is well approximated by a single mass scale, $M_{\tilde Q}$,
and allow for a large hierarchy between the squark and slepton
masses. We include only (s)quark loops in the calculation, because the
correction is enhanced by a color factor and the number of
generations. The remaining corrections are small, and if we did
include them we expect our conclusions would not change.

In the next section we discuss the radiative correction calculation.
We point out that, to a good approximation, the one-loop $t$-channel
amplitudes can be rewritten in the same form as the tree-level
amplitudes, with the replacement of the tree-level parameters with
renormalized effective parameters. Hence we introduce the effective
coupling, the effective masses, and the effective mixing matrix.  In
section 3 we show our numerical results. We show the renormalization
scale and squark mass dependence of the various production modes, for
the different electron polarizations. We compare the tree-level,
leading-log, effective, and full cross-section calculations. The
results show that the effective theory approximation describes the
full result very well. We also discuss the $A$-term dependence.

In section 4 we discuss how well we can measure the squark loop
correction to the coupling, and thereby constrain the squark mass,
assuming both slepton and chargino production are possible.  We show
the statistical significance of the results by combining our knowledge
of the superpartner masses and cross-sections.  The slepton production
cross-section is a sensitive function of the slepton mass.  The
uncertainty in the slepton mass measurement is quite important in this
analysis. We emphasize that the chargino production cross-section does
not have this problem of the final-state mass uncertainty.  In the
last section, section 5, we give our conclusions.

\section{The one-loop cross-section}

In this section we discuss the calculation of the cross-section of
$e^-e^+\rightarrow\tilde\ell_i\tilde\ell_j^*$, where
$\tilde\ell_i=(\tilde e_L^-,\tilde e_R^-,\tilde\nu_e)$, including
one-loop (s)quark corrections. The full result is explicitly given in
Appendix A; here we restrict ourselves to outline the general features
of the calculation.

As a specific example, we consider the process $e^-e^+\rightarrow\tilde
e_R\tilde e_R^*$.  The tree-level amplitude is given by
\begin{equation}
{\cal M}_{RR}^{(0)} = 2\bar v\psla
\biggl[ {e^2\over s}
+ {g_Y^2\over s-M_Z^2}\left(s_W^2 - {1\over2}P_L\right)
- \sum_{i=1}^4{g_Y^2 N_{i1}^*N_{i1}
\over p^2 - m_{\tilde\chi^0_i}^2}P_R\biggr]u\ ,
\end{equation}
where $s$ is the center-of-mass energy, $p$ is the $t$-channel
momentum, and $P_{L,R}=(1\mp\gamma_5)/2$. The sine of the weak mixing
angle is denoted $s_W$, $N_{ij}$ is the neutralino mixing matrix, and
the $e^\pm$ wave functions are denoted $\bar v$ and $u$.

To evaluate the one-loop amplitude, we treat all the parameters
appearing in this tree-level expression as running \DR~ (dimensional
reduction with modified minimal subtraction \cite{DR}) quantities, and
add the contributions from the one-loop diagrams (see Fig.~1).  Note
that the (s)quark loop corrections do not give rise to external
wave-function renormalization.

We take the gauge boson pole masses and the standard model \MS~
electromagnetic coupling $\hat\alpha_{\rm SM}(M_Z)$ as inputs.  The
\DR~ parameters are related to the corresponding input parameters by
familiar expressions. (Hatted symbols are \DR~ renormalized quantities.
We do not hat the \DR~ soft supersymmetry breaking parameters.)  For
example, the running gauge boson masses are given by
\begin{equation}
{\hat M_W^2}(Q) = M_W^2 + {\cal R}e\,\hat\Pi^T_{WW}(M_W^2)\ ,\;\; 
{\hat M_Z^2}(Q) = M_Z^2 + {\cal R}e\,\hat\Pi^T_{ZZ}(M_Z^2)\ ,  \label{eq6}
\end{equation}
where the $\hat\Pi^T$ are the transverse parts of the \DR~ renormalized
gauge-boson self-energies.  The gauge coupling in the full theory
$\hat e(Q)$ is related to that of the effective theory 
(i.e. the standard model in the \MS~ scheme) as
\footnote{The constant term $(e^2/24\pi^2)$ for \MS~ to \DR~ conversion
is omitted because we systematically ignore the gauge boson loops.}
\begin{equation}
\hat e(Q) = \hat e_{\rm SM}(Q)\left(1 
+\frac{\hat e_{\rm SM}^2 N_c}{48\pi^2}
\sum_{\tilde{q}_i}Q_{\tilde{q}}^2
\ln\frac{Q}{m_{\tilde{q}_i}}\right)\ . \label{eq3}
\end{equation}
The SU(2)$_L$ and U(1)$_Y$ gauge couplings are then obtained in terms
of the gauge boson running masses,
\begin{eqnarray}
\hat g_Y^2(Q)&=&\hat e^2(Q)\frac{\hat M_Z^2}{\hat M_W^2}
\ ,\label{eq4}\\
\hat g_2^2(Q)&=&\hat e^2(Q)\frac{\hat M_Z^2}{\hat M_Z^2-\hat M_W^2}
\ .\label{eq5}
\end{eqnarray}
Note that the argument of the gauge-boson self-energies is the
external momentum squared.  They implicitly depend on the
renormalization scale $Q$. Explicit expressions for
$\hat\Pi^T_{WW,ZZ}$ are given in \cite{GB,P}.  Since we only include
the (s)quark loop corrections, for consistency the \DR~ parameters only
run by virtue of the (s)quark contributions to their RGE's.

When $M_1$, $M_2$, $\mu$, $\tan\beta$ are needed, we use $M_1(M_1)$,
$M_2(M_2)$, $\mu(|\mu|)$, $\tan\beta(M_Z)$.  The scale choice of each
parameters has been made because the relations
$m_{\tilde{\chi}^0_1}\simeq M_1(M_1)$, $m_{\tilde{\chi}^0_2} \simeq
M_2(M_2)$, $m_{\tilde{\chi}^0_3} \simeq |\mu|$ hold when ${\rm
max}(M_1,M_2,|\mu|)\gg M_Z$, and $M_{\tilde{q}_i}\lsim {\rm
min}(M_1,M_2,|\mu|)$.

With these counter-terms and the one-loop diagram contributions
presented in Appendix A, we obtain the full one-loop amplitude.  The
differential  cross-section is then computed,
\begin{equation}
{d\sigma\over dt} = {1\over16\pi s^2} \left|{\cal M}\right| ^2\ .
\end{equation}
To obtain the total cross-section, we integrate over the 
$t$-channel momentum numerically.  The
cross-sections for the other selectron and sneutrino production
processes are obtained similarly.

In Appendix B, we show how the one-loop corrected $t$-channel
amplitude can be well approximated by a tree-level form,
\begin{equation}
{\cal M}^t_{RR} = 2\bar
v\psla\biggl[-\sum_{i=1}^4\frac{\bar{g}^2_{e\tilde{e}_R\tilde{B}}
\overline{N}_{i1}^*\overline{N}_{i1}(p^2) }
{p^2-\overline{m}_i^2(p^2)}P_R\biggr]u\ , \label{eff amp}
\end{equation}
where $\bar{g}_{e\tilde{e}_R \tilde{B}}(p^2)$ 
is the effective bino coupling defined as 
\begin{equation}
\bar{g}_{e\tilde{e}_R \tilde{B}}(p^2)=\hat g_Y(Q^2)
\biggl(1-\frac{1}{2}\tilde{\Sigma}^L_{11}(p^2)\biggr)\ ,
\end{equation}
and $\tilde{\Sigma}^L_{11}(p^2)$ is the bino-bino component of the
neutralino two-point function defined in Eq.~(\ref{eqb4}). The
$\overline{N}_{ij}$ and $\overline{m}_i$ are the effective neutralino
mixing matrix and neutralino masses obtained by diagonalizing the
effective neutralino mass matrix $\overline{Y_{ij}}$, defined in
Eq.~(\ref{eqb6}).  We note that $\overline{m}_i(\overline{m}_i^2)$
are the chargino and neutralino pole masses.  Also, $\bar{g}_i$,
$\overline{N}_{ij}$, and $\overline{m}_i$ are scale independent
quantities to ${\cal O}(\alpha)$. For example, the scale dependence of
$\hat g_Y(Q)$ is cancelled by the implicit scale dependence of
$\tilde{\Sigma}^L_{11}(p^2)$.

The physical correction appears as a violation of the tree-level
relation $\bar g_{\ell\tilde\ell\tilde\chi}=\hat g^{\rm eff}$, where
$\hat g^{\rm eff}$ is the effective coupling in the standard model. It
is obtained by Eqs.~(3-6), including only quark contributions in the
self-energies. The $\log M_{\tilde Q}$ terms in $\bar
g_{\ell\tilde\ell\tilde\chi} - \hat g$ and $\hat g^{\rm eff} - \hat g$
correspond to the change in the beta functions above and below the
squark threshold.  Also, the leading logarithms of the corrections
$\delta g/g = ( \bar g_{\ell\tilde\ell\tilde\chi}-\hat g^{\rm
eff})/\hat g^{\rm eff}$ at $Q=m_{\tilde\ell}$ are exactly those of
Eq.~1, showing that the RGE approach of Ref.~\cite{C} gives the proper
results.

The other effective couplings $\bar{g}_{e\tilde{e}_L\tilde{B}}$,
$\bar{g}_{e\tilde{e}_L\tilde{W}}$ and
$\bar{g}_{e\tilde{\nu}_e\tilde{W}}$ are given by analogous formulae
(see Eq.~(\ref{eqb10})). The expressions for
$\bar{g}_{e\tilde{e}_L\tilde{W}}$ and
$\bar{g}_{e\tilde{\nu}_e\tilde{W}}$ are different. However, they
approach each other in the large $M_{\tilde{Q}}$ limit.  The
difference between these effective couplings arises from finite
electroweak symmetry breaking contributions proportional to the vacuum
expectation value of the Higgs bosons. By gauge symmetry, the leading
$\log M_{\tilde{Q}}$ correction is the same for the different
effective couplings. In section 3, we will see that the various finite
corrections are not large compared to the leading $\log M_{\tilde{Q}}$
correction.

In contrast to the $t$-channel amplitudes, the squark corrections to
the $s$-channel amplitudes are proportional to $(M_Z^2\ {\rm or}
\ m_t^2)/M_{\tilde Q}^2$.  Since we take as inputs the standard model
coupling, and the $W$- and $Z$-boson pole masses, the decoupling
theorem applies, and the effect of the heavy squarks is negligible in
the large squark mass limit.

Part of the radiative corrections we are considering are closely
related to the corrections to the neutralino and chargino masses
\cite{P,ino masses}.  Since the chargino and neutralino masses will be
measured in future collider experiments, we will in some of the
discussion take the pole masses $m_{\tilde{\chi}^0_1}$,
$m_{\tilde{\chi}^+_1}$, $m_{\tilde{\chi}^0_3}$ as inputs.

We define $M_1^{\rm eff}$, $M_2^{\rm eff}$, $-\mu^{\rm eff}$ as the
(1,1), (2,2), and (3,4) components of the effective neutralino mass
matrix $\overline{Y}_N$. $M_2^{\rm eff}$ and $\mu^{\rm eff}$ could
also be defined as the (1,1), (2,2) components of the effective
chargino mass matrix $\overline{Y}_C$. These definitions differ by
${\cal O}(M_Z^2/{\rm max}(M_1^2,M_2^2,\mu^2))$, as do the different
definitions of the wino effective couplings.  The grand unification
condition leads to the one-loop relations
$M_1/\alpha_1=M_2/\alpha_2=M_3/\alpha_3$.  We comment that there is a
$\ln M_{\tilde{Q}}$ correction to the $M_i^{\rm eff}/\alpha_i^{\rm
eff}$ ratios. We set $M_2\sim2M_1$, but we do not impose any such
relation in our analysis.

\section{Numerical results}

We next study the numerical dependence of the one-loop corrected 
cross-sections of $e^-e^+\rightarrow\tilde\ell_i\tilde\ell_j^*$
($\tilde\ell_i=(\tilde e_L^-,\tilde e_R^-,\tilde\nu_e)$) on the squark
mass.  We consider the case where the initial electron is completely
longitudinally polarized.  We therefore treat the following eight
modes,
\begin{eqnarray}
e^-_Le^+& \rightarrow& 
\tilde{e}_L^-\tilde{e}_L^+\ , \;
\tilde{e}_R^-\tilde{e}_R^+\ , \;
\tilde{e}_L^-\tilde{e}_R^+\ , \;
\tilde{\nu}_e\tilde{\nu}_e^*\ , \nonumber \\
e^-_Re^+& \rightarrow& 
\tilde{e}_L^-\tilde{e}_L^+\ , \;
\tilde{e}_R^-\tilde{e}_R^+\ , \;
\tilde{e}_R^-\tilde{e}_L^+\ , \;
\tilde{\nu}_e\tilde{\nu}_e^*\ .  \label{eq10}
\end{eqnarray}
In the squark sector, we set
\begin{equation}
M_{\tilde{Q}}=M_{\tilde{U}}=M_{\tilde{D}}\ ,\;\;
A\equiv A_t/M_{\tilde{Q}}=A_b/M_{\tilde{Q}}\ , 
\end{equation}
where $M_{\tilde{Q}},\ M_{\tilde{U}}$, and $M_{\tilde{D}}$ are the
soft-breaking masses of the left-handed, right-handed up, and
right-handed down-type squarks, respectively, and $A_t$, $A_b$ are the
soft-breaking Higgs-squark-squark trilinear couplings, in the
convention of Ref.~\cite{GH}. We assume generation independence for
these parameters.

The renormalization scale independence of the one-loop cross-section
serves as an important check of the calculation.  In Fig.~2 we show
the $Q$ dependence of the tree-level cross-sections (defined as the
first terms of Eqs.~(\ref{eqa2}--\ref{eqa6})) and the corrected ones,
for $M_1(M_1)=100$~GeV, $M_2(M_2)=200$~GeV, $\mu(|\mu|)=-300$~GeV,
$\tan\beta(M_Z)=4$, $m_{\tilde{\ell}}=200$~GeV,
$M_{\tilde{Q}}=1000$~GeV, and $\sqrt{s}=500$~GeV.  The tree-level
cross-sections vary linearly with $\log Q$. This dependence is due to
the running of $\hat g_i(Q),\ \hat M_Z,$ and $M_i(Q)$.  One can see
that this $Q$ dependence almost vanishes after the inclusion of the
one-loop contributions. The remaining $Q$ dependence of the corrected
cross-sections, which comes from residual higher-order corrections in
the numerical calculation, is very small for $M_Z<Q<1$~TeV.  The
renormalization scale dependence suggests the uncertainty in the
calculation due to higher order corrections.

We next consider the squark mass dependence of the corrected
cross-sections.  As stated, we take the standard model electromagnetic
coupling and the $W$ and $Z$ boson pole masses as inputs.  Here we
also take the three pole masses $(m_{\tilde{\chi}^0_1}$,
$m_{\tilde{\chi}^+_1}$, $m_{\tilde{\chi}^0_3}$), and $\tan\beta(M_Z)$
as inputs.  We assume $|\mu|\gg M_Z$, in which case $M^{\rm
eff}_1\simeq m_{\tilde{\chi}^0_1}$, $M^{\rm eff}_2\simeq
m_{\tilde{\chi}^+_1}$, and $|\mu^{\rm eff }|\simeq
m_{\tilde{\chi}^0_3}$ hold, where $M^{\rm eff}_1$, $M^{\rm eff}_2$,
and $-\mu^{\rm eff}$ are the (1,1), (2,2), and (3,4) elements of the
effective neutralino mass matrix $\overline{Y}_{ij}$, defined in
Eq.~(\ref{eqb6}).

We show in Fig.~3 the $M_{\tilde Q}$ dependence of the cross-sections
for $m_{\tilde{\chi}^0_1}=100$~GeV, $m_{\tilde{\chi}^+_1}=200$~GeV,
$m_{ \tilde{\chi}^0_3}=300$~GeV, $\tan\beta(M_Z)=4$,
$m_{\tilde{\ell}}=200$~GeV, $A=0$, $\mu<0$, and $\sqrt{s}=500$~GeV.
Here we normalize the cross-sections to the tree-level values.  The
tree-level cross-sections are functions of the standard model gauge
couplings $\hat g_i^{\rm eff}(Q)$ (at $Q=200$ GeV) and the tree-level
$M_i,\ \mu$ and $\tan\beta$. The latter parameters are found by
inverting the tree-level relations between $M_i, \mu, \tan\beta$ and
the input masses $m_{\tilde\chi_1^0},\ m_{\tilde\chi_1^+}$ and
$m_{\tilde\chi_3^0}$. Hence, the tree-level cross-sections are
independent of the squark mass scale.  The numerical values of the
tree-level cross-sections are listed in Table~1.

\begin{table}[t]
\begin{center} 
\begin{tabular}{|l|c|c|c|c|c|}
\hline
& $\tilde{e}^-_L\tilde{e}^+_L$ & $\tilde{e}_R^-\tilde{e}_R^+$ & 
$\tilde{e}_L^-\tilde{e}_R^+$ & $\tilde{e}_R^-\tilde{e}_L^+$ & 
$\tilde{\nu}_e\tilde{\nu}_e^*$ \\ 
\hline
$e^-_L$ & 105 & 8.83 & 121 & 0 & 678 \\
\hline
$e^-_R$ & 8.83 & 235 & 0 & 121 & 9.81 \\
\hline
\end{tabular}
\caption{Total tree-level cross-sections, in fb, of 
$e^-e^+\rightarrow\tilde\ell_i\tilde\ell_j^*$ 
for the various modes. Input parameters are 
given in the text.}
\end{center}
\end{table}

The one-loop corrected cross-sections of the modes which have a
$t$-channel contribution are similar to tree-level ones at
$M_{\tilde{Q}}\lsim 300$~GeV, but increase linearly with $\log
M_{\tilde{Q}}$.  Because the effective masses are equal to the input
pole masses to ${\cal O}((p^2-m_{\tilde\chi}^2)/M_{\tilde Q}^2)$, the
squark mass dependence of the one-loop corrected cross-sections is
primarily due to the difference between the effective theory gauge and
gaugino couplings.

The two channels which have $\tilde{W}$ contributions
($e^-_Le^+\rightarrow\tilde{e}_L^-\tilde{e}_L^+
,\tilde{\nu}_e\tilde{\nu}_e^*$) show the largest $M_{\tilde{Q}}$
dependence.  The destructive interference between the $s$- and
$t$-channel amplitudes accounts for the enhancement of the $\tilde
e_L^-\tilde e_L^+$ correction relative to the
$\tilde{\nu}_e\tilde{\nu}_e^*$ case.  Other channels
$e^-_Le^+\rightarrow\tilde{e}_L^-\tilde{e}_R^+$,
$e^-_Re^+\rightarrow\tilde{e}_R^-\tilde{e}_R^+,
\tilde{e}_R^-\tilde{e}_L^+$ show smaller $M_{\tilde{Q}}$ dependence
from $\tilde{B}$ contributions. Nevertheless, these $M_{\tilde{Q}}$
dependences are significantly larger than the renormalization scale
dependence of the corrected cross-sections (see Fig.~2). In contrast, the
remaining channels, which have only $s$-channel contributions, show very
little $M_{\tilde{Q}}$ dependence, as explained in Section 2.

We now consider two approximations to the full calculation.  First, we
consider the effective theory approximation (ETA), where the
effective masses, couplings, and mixing matrices are used in the
$t$-channel amplitudes (see Eq.~(\ref{eff amp})). We make a further
approximation in the ETA by fixing the momenta of the effective
parameters to an average $t$-channel momentum, $p^2=-m_{\tilde\ell}^2$.
The second approximation is the leading log approximation (LLA), which
is the ETA in leading logarithm approximation.

The full calculation differs from the ETA by ${\cal O}(M_Z^2/M_{\tilde
Q}^2)$ terms which arise when factorizing the amplitude (and by ${\cal
O}((p^2+m_{\tilde\ell}^2)/M_{\tilde Q}^2)$ corrections due to fixing
$p^2$).  The LLA differs additionally by constant and ${\cal
O}(1/M_{\tilde Q}^2)$ corrections. Figs.~4(a) and (b) shows the full
correction, the ETA, and the LLA, for
$\sigma(e^-e^+\rightarrow\tilde{e}^-_R\tilde{e}^+_R)$ and
$\sigma(e^-e^+\rightarrow$ $\tilde{\nu}_e\tilde{\nu}_e^*)$,
respectively.  In Fig.~4(b) we use the effective coupling
$\bar{g}_{e\tilde{e}_L\tilde{W}}(-m_{\tilde{\ell}}^2)$, rather than
$\bar{g}_{e\tilde{\nu}_e\tilde{W}}(-m^2_{\tilde{\ell}})$, which
introduces additional ${\cal O}((M_Z^2,\,m_t^2)/M_{\tilde Q}^2)$
differences between the full and ETA calculations. Figs.~4 show that
the full calculation is well approximated by the ETA, even for small
squark masses. The LLA does correctly describe the logarithmic
increase in the full correction (as the name implies), but there is a
constant difference of 0.6\% [1.0\%] seen between the full and LLA
curves in Fig.~4(a)~[(b)]. This constant difference is largely
accounted for by the correct matching of the effective coupling,
where, at zero momentum, and for zero quark mass, the logarithms in
Eq.~(\ref{ln msq}) are replaced as \cite{CFPa}
$$\ln\left({M_{\tilde Q}\over m_{\tilde\ell}}\right) \qquad
\longrightarrow \qquad \ln\left({M_{\tilde Q}\over
m_{\tilde\ell}}\right) - 3/8\ .$$

In Fig.~4 there is a slight discrepancy between the ETA and the full
results at very large $M_{\tilde Q}$. This is due to the resummation
of the effective mass corrections inherent in the derivation of the
effective amplitude.

Before leaving this section, we comment on the dependence of the
cross-sections on the parameter
$A=A_t/M_{\tilde{Q}}=A_b/M_{\tilde{Q}}$, which can give sub-leading
corrections through the left-right mixing of third generation squarks.
One might expect that the $A$ dependence could become significant for
small ($\sim 1$) values of $\tan\beta$.  Fig.~5 shows that for
$\tan\beta(M_Z)=1.5$ the dependence is much smaller than the
$M_{\tilde{Q}}$ dependence.  For the case $\mu<0$, the cross-section
is almost independent of $A$ if $M_{\tilde Q}\gsim1$ TeV. For $\mu>0$,
there is more dependence. For example, for $M_{\tilde Q}=1$ TeV, the
cross-section increases by about 3 per cent as $A$ varies from $-3$ to
$+3$.

\section{$\Delta\chi^2$ analysis}

In the previous section, we have shown the squark mass dependence of
the slepton production cross-section $\sigma^{\rm one-loop}(e^-e^+
\rightarrow \tilde{\ell}\tilde{\ell}^*)$. We found the cross-sections
are well described by that of an effective theory, where the
tree-level relations between the fermion-sfermion-gaugino couplings
and the gauge couplings are renormalized. In this section, we consider the
determination of the squark mass $M_{\tilde{Q}}$ through the
measurement of the slepton production cross-section.

As described in section 2, the sfermion and chargino/neutralino
production cross-sections depend on the gaugino mass parameters $M_1$
and $M_2$, the Higgsino mass parameter $\mu$, the ratio of the vacuum
expectation values $\tan\beta$, and the sfermion mass
$m_{\tilde\ell}$.  The cross-sections also depend on the effective
fermion-sfermion-gaugino coupling $\bar{g}_{f\tilde{f}\tilde\chi}$,
which is nicely parameterized by $\log M_{\tilde Q}$.  By constraining
$\bar{g}_{f \tilde{f}\tilde\chi}$ we can determine $M_{\tilde Q}$, if
the rest of the parameters are known accurately enough.

It has been demonstrated that an accurate determination of
$m_{\tilde{\chi}}$ and $m_{\tilde\ell}$ is indeed possible if
sfermions $\tilde\ell$ are produced and dominantly decay into a
charged lepton $\ell'$ and a chargino or neutralino $\tilde\chi$. The
simple kinematics of the two body decays result in a flat energy
distribution of the final state leptons between the endpoint energies
$E_{\rm min}^{\ell'}$ and $E_{\rm max}^{\ell'}$, which are simple
functions of $\sqrt{s}$, $m_{\tilde\ell}$, and $m_{\tilde{\chi}}$.
The measured endpoint energies therefore determine $m_{\tilde\ell}$
and $m_{\tilde\chi}$.

In Ref. \cite{TSUKA}, the results of detailed MC simulations were
presented for $\tilde{\mu}\rightarrow\mu\tilde{\chi}^0_1$ and
$\tilde{e}\rightarrow e\tilde{\chi}^0_1$.  It was shown that the
$\tilde{\ell}_R$ and $\tilde{\chi}^0_1$ masses could be measured to
better than 1\%.  Recently, Baer et al. \cite{BMT} performed a MC
study for the case that left-handed sfermions are produced and decay
into a gaugino-like chargino or neutralinos. In their example called
point 3, $\tilde{\nu}_e\tilde{\nu}_e^*$ production is followed by
$\tilde{\nu}_e^{(*)}\rightarrow e^\mp\tilde{\chi}^{\pm}_1$.  The decay
mode $\tilde{\nu}_e\tilde{\nu_e}^*$ $\rightarrow
e^-e^+\tilde{\chi}_1^+\tilde{\chi}^-_1$ $\rightarrow e^-e^+ \mu 2j$
($\nu_{\mu}$ $2\tilde{\chi}^0_1)$ is background free and the
measured electron endpoint energies allow for a 1\% measurement of
$m_{\tilde{\chi}^{\pm}_1}$ and $m_{\tilde\nu}$\footnote{Notice that
the mass errors are considerably improved over those obtained from the
$e^+e^-\rightarrow \tilde{\chi}^+_1\tilde{\chi}^-_1$ process
\cite{TSUKA}. The mass measurement from the jet energy distributions
in chargino decay $\tilde{\chi}^+_1\rightarrow\tilde{\chi}^0_1
q\bar{q}$ gives only a few percent measurement of the masses due to
the uncertainty in the jet momentum.}.

The results of Ref. \cite{BMT} encourage us to consider their example
point 3. The chosen parameter set corresponds to $m_{\tilde{\nu}_e}=
207$ GeV, $m_{\tilde{\chi}^+_1}=96$ GeV, $m_{\tilde{\chi}^0_1}=45$~GeV
and $m_{\tilde{\chi}^0_3}=270$~GeV, and the lightest chargino and
neutralinos are gaugino-like. Their study suggests that we can take
$m_{\tilde{\chi}^+_1},$ $m_{\tilde{\chi}^0_1}$, and
$m_{\tilde{\nu}_e}$ as well constrained input parameters. For 20
fb$^{-1}$ of luminosity, their MC simulations show that at 68\% CL,
$(\delta m_{\tilde{\chi}^+_1},\ \delta m_{\tilde{\nu}_e}$) = (1.5~GeV,
2.5~GeV).

In the following we estimate the statistical significance of the
radiative correction to the production cross-section. To most
effectively constrain the squark mass we would take into account all
the possible channels.  However, for the purposes of this paper, it
suffices for us to focus solely on the $\tilde{\nu}_e$ production
cross-section, because it is larger than 1 pb for a left-handed
electron beam, and larger than the other sparticle production
cross-sections at $\sqrt{s}=500$~GeV.

The sensitivity of the sneutrino production cross-section on the
chargino mixing is determined by the sign of $\mu$. For $\mu<0$, the
gaugino/Higgsino mixing is suppressed, and hence the cross-section
depends mildly on $\tan\beta$. Also, there is very little
$m_{\tilde\chi_3^0}$ dependence in this case. For $\mu>0$, the mixing
gives rise to significant $\tan\beta$ and $m_{\tilde\chi_3^0}$
dependence.  The heavy chargino and neutralinos are produced once
$\sqrt{s}>2|\mu|$, and we assume that these masses can be constrained
to a reasonable range\footnote{The heavier chargino and neutralinos
dominantly decay to a $W$ or $Z$ and a wino-like neutralino or
chargino (through its suppressed Higgsino component).  The dominant
decay process gives maximally 8 jets in the final state, and the
background to the process is small.  If the energy distribution of the
gauge bosons in these decays could be measured, the endpoint technique
could also be applied here.}. Also, if $\mu$ is not too large, the
splitting of the heavy chargino and neutralino masses is quite
different for the different signs of $\mu$.  If the splitting can be
measured, the sign of $\mu$ might be determined. For example, at point
3 with $m_{\tilde\chi_3^0}=270$ GeV and $\mu<0$ ($\mu>0$),
$m_{\tilde\chi_4^0}-m_{\tilde\chi_3^0} = 5$ (27) GeV. The reasonably
large production cross-sections of the heavier chargino and
neutralinos (${\cal O}(100)$ fb at $\sqrt{s}=800$ GeV) give us hope
that these mass differences can be distinguished. As long as
$m_{\tilde\chi_3^0}$ is known to ${\cal O}(5$-$10\%)$ and the sign of
$\mu$ can be determined, the chargino mixing uncertainty will not
significantly affect the determination of the effective coupling
$\bar{g}_{e\tilde{\nu}_e\tilde{W}}$.

We would first like to provide a feel for the sensitivity to the
squark mass scale and $\tan\beta$ in the ideal case where we ignore
the slepton and gaugino mass uncertainties.  In Fig.~6 we show the
statistical significance of the loop correction by plotting
contours of constant cross-section. Here we fix the sneutrino mass and
determine $\mu,\ M_1$ and $M_2$ by fixing the one-loop corrected
masses $m_{\tilde{\chi}^0_1}$, $m_{\tilde{\chi}^0_3}$, and
$m_{\tilde{\chi}^+_1}$.  Rather than plot the contours according to
the value of the production cross-section, we plot the contours
corresponding to the number of standard deviations of the
fluctuation of the accepted number of events.  The 1-$\sigma$
fluctuation corresponds to $\sqrt{N_{\rm input}}$, where $N_{\rm
input}$ is our nominal value of the number of events at
$M_{\tilde{Q}}=1000$~GeV and $\tan\beta(M_Z)=4$. The accepted number
of events $N$ is given by
\begin{equation}
N=A\cdot\sigma(e^-_Le^+\rightarrow \tilde{\nu}_e\tilde{\nu}_e)\times  
\left( {\rm BR} (\tilde{\nu}_e\rightarrow e\tilde{\chi}^+_1)
\right)^2 \times 100\ {\rm fb}^{-1}. 
\end{equation}
Here we took ${\rm BR}(\tilde{\nu}_e\rightarrow e \tilde{\chi}^+_1) =
0.6$ and overall acceptance $A=0.28$ \footnote{The acceptance given in
Ref.~\cite{TSUKA} is 45\% for $m_{\tilde{e}_R}=142$~GeV,
$m_{\tilde{\chi}^0_1}=118$~GeV and $\sqrt{s}=350$~GeV.  We further
scale our acceptance down, because our calculation of cross-section
does not include the effects of initial state radiation, beam energy
spread, and beamstrahlung.}.  The number of accepted events at our
nominal point $N_{\rm input}$ is about 12800 for $\mu<0$ and
11900 for $\mu>0$.

We show the contours for $\mu<0$ in Fig.~6(a).  If $\tan\beta$ is well
measured, $M_{\tilde Q}$ is constrained to
$M_{\tilde{Q}}=1000^{+370}_{-280}$~GeV at 1-$\sigma$ significance.  If
instead we assume the constraint $2<\tan\beta<8$, the mild $\tan\beta$
dependence yields $700 < M_{\tilde{Q}} < 1900$ GeV. In the case
$\mu>0$ (Fig.~6(b)), increasing the squark mass can be compensated for
by decreasing $\tan\beta$, and measuring sneutrino production then
determines a region of the $(M_{\tilde Q}, \ \tan\beta)$
plane. As with $\mu<0$, $M_{\tilde Q}$ is decently constrained if
$\tan\beta$ is known ($M_{\tilde{Q}} =
1000^{+350}_{-450}$~GeV).  However, if we have $2<\tan\beta<8$, only
the uninteresting bound $M_{\tilde Q}<10$ TeV can be achieved.

In Fig.~6(c) we show the contours again for $\mu>0$, but now we raise
$m_{\tilde\chi_3^0}$ to 500 GeV. Larger values of $\mu$ lead to
weaker $\tan\beta$ dependence. However, in this case the cross-section
is still sensitive to the heavy chargino mass $m_{\chi_2^+}$.  If the
heavier chargino is beyond the kinematical reach of the collider,
there may be considerable uncertainty in the predicted
cross-section. The cross-section is pretty much independent of $\mu$
once $\mu>850$ GeV, in this case. If we vary $\mu$ from 450 to 850 GeV
the cross-section varies by about 2\%. It is good to keep in mind that
$\mu$ may be constrained in a situation such as this by measuring the
left-right asymmetry of the lighter chargino production cross-section.

Figs.~6 illustrate that we might have to rely on independent
measurements of $\tan\beta$ in order to obtain a nice constraint on
$M_{\tilde Q}$.  For many regions in parameter space we expect
$\tan\beta$ will indeed be known fairly well. For example, it was
pointed out by Ref. \cite{FPMT} that the measurement of the chargino
forward-backward asymmetry provides a strong constraint on $\tan\beta$
if $M_2\sim|\mu|$. Also, for not too large slepton masses and
$\tan\beta\sim1$, the measurement of the mass difference between
$\tilde{e}_L$ and $\tilde{\nu}_e$ gives a sensitive measurement of
$\tan\beta$.  Other measurements can be instrumental in pinning down
$\tan\beta$. For example, the Higgs sector constraints on $\tan\beta$
are considered in Ref.~\cite{FM}, which tend to give decent bounds
for $\tan\beta<10$.

We now turn to the effect of the mass uncertainties. The sneutrino
production cross-section depends on the masses $m_{\tilde{\nu}_e}$,
$m_{\tilde{\chi}^0_1}$, $m_{\tilde{\chi}^0_3}$ and
$m_{\tilde{\chi}^+_1}$.  The uncertainties in these masses will of
course worsen our ability to constrain the squark mass scale.  Of
these massses the cross-section is most sensitive to the sneutrino
mass.  All of the same chirality scalar production cross-sections
suffer from the strong $\beta_{\tilde\ell}^3$ kinematic dependence
($\beta_{\tilde\ell} = \sqrt {1 - (m_{\tilde{\ell}}/E_{\rm
beam})^2}$)\footnote{The production cross-section
$\sigma(e^-e^+\rightarrow \tilde{e}_R\tilde{e}_L)$ is proportional to
$\beta_{\tilde\ell}$, due to the chiral structure of
$\ell$-$\tilde{\ell}_{L(R)}$-$\tilde{\chi}^0_i$ coupling.}. Near
threshold this results in an especially large sensitivity to the
final-state mass.  For example, a 1\% uncertainty of $m_{\tilde\ell}$
results in $\delta\beta_{\tilde\ell}/\beta_{\tilde\ell}=1.8\%$, for
$m_{\tilde{\ell}}=200$~GeV, and $\sqrt{s}=500$~GeV. This immediately
translates into a 5.3\% error in the total cross-section.  Although a
simple statistical scale-up of the results of Ref.~\cite{BMT} implies
a sneutrino mass uncertainty of only 0.3\%, this nevertheless leads to
a significant degradation in our ability to constrain the squark mass
scale. (Systematic errors might be the limiting factor here.)

Note, however, that the measurement of the sneutrino mass in
Ref.~\cite{BMT} was obtained by studying a small fraction of the total
sneutrino decay modes.  The mode they studied,
$e^-e^+\rightarrow\tilde\nu_e\tilde\nu_e^*$ $\rightarrow e^-e^+\mu 2j
(\nu_{\mu} 2\tilde{\chi}^0_1)$, amounts to only about 4\% of the total
sneutrino decays. Using other modes, such as $e^-e^+4j
(2\tilde{\chi}^0_1)$, might reduce the mass error even
further. Because the $\tilde{\nu}_e$ production cross-section is
significantly larger than the other slepton cross-sections, isolating
the various sneutrino signatures is less affected by SUSY backgrounds
such as $e^-e^+\rightarrow \tilde{e}^-_L\tilde{e}^+_L$ $\rightarrow
e^-e^+ 4j(2\tilde{\chi}^0_1)$. In this discussion we assume the errors
are dominantly statistical.

Now we show the constraint on the squark mass $M_{\tilde{Q}}$ after
taking into account the uncertainty of the masses $\delta
m_{\tilde{\nu}_e}$ and $\delta m_{\tilde{\chi}^+_1}$.  For this
purpose, we define a $\Delta\chi^2$ function as
\begin{eqnarray}
\Delta\chi^2&=&\frac
{\left( N( m_{\tilde{\chi}^0_1}, m_{\tilde{\chi}^0_3},
m_{\tilde{\chi}^+_1}, \tan\beta, m_{\tilde{\nu}_e}, M_{\tilde{Q}})
- N_{\rm input}\right)^2}
{N_{\rm input}}\nonumber
\\
&
+&\frac{(m_{\tilde{\chi}^+_1} -m_{\tilde{\chi}^+_{1 \ {\rm input}}} )^2}
{(\delta m_{\tilde{\chi}^+_1})^2}
+
\frac{\left(m_{\tilde{\nu}_e} -m_{\tilde{\nu}_{e \ \rm input}} \right)^2}
{(\delta m_{\tilde{\nu}_e})^2}\ .
\end{eqnarray}
Here we consider the $\mu<0$ case, in which the $\tilde\chi_1^0$ and
$\tilde\chi_3^0$ mass dependencies are negligible. We also assume that
$\tan\beta$ is well measured.  In Fig.~7 we plot $\Delta\chi^2_{\rm
min}$ against $M_{\tilde{Q}}$, where $\Delta\chi^2_{\rm min}$ is a
minimum of $\Delta\chi^2$ with respect to variations in
$m_{\tilde{\chi}^+_1}$ and $m_{\tilde{\nu}_e}$. By construction, the
region of $M_{\tilde Q}$ where $\sqrt{\Delta\chi^2_{\rm min}}<1,2,...$
corresponds to $1,2,...$-$\sigma$ error of the squark mass when the
chargino and sneutrino mass uncertainties are taken into account.  The
sneutrino mass uncertainty reduces the sensitivity of the production
cross-section to $M_{\tilde{Q}}$ considerably, because the effect of
increasing $M_{\tilde{Q}}$ can be compensated for by a small
increase in $m_{\tilde{\nu}_e}$.  On the other hand, we do not find
any significant effect due to non-zero $\delta m_{\tilde{\chi}^+_1}$.

{}From Fig.~7 we see that in this case, even with the sneutrino mass
uncertainty, we can reasonably constrain the squark mass scale. For
example, at the 1-$\sigma$ level with $M_{\tilde Q}=1$ TeV, we
constrain $M_{\tilde Q}$ to $1^{+1.2}_{-0.5}$ TeV, using the naive
scale up (from 20 fb$^{-1}$ to 100 fb$^{-1}$) of the statistical
errors of Ref.~\cite{BMT}.  This corresponds to the difference between
the gauge and gaugino effective couplings, $\delta g_2/g_2 =
0.011\pm0.006$. This can be compared to the estimate of the constraint
$\delta g_2/g_2 = \pm0.02$ from the chargino production measurement
\cite{CFPb}. Such comparisons are sensitive to different choices of
parameter space and other assumptions.  The sensitivity of the
constraint on the mass errors can be seen by changing the errors. If
we reduce the mass uncertainties by a factor of 2, we find the
interesting constraint $600 <M_{\tilde{Q}}< 1500$ GeV.

Fig.~7 illustrates the large dependence on the mass uncertainties.
This sensitivity is greatly reduced in the case of chargino
production. As we mentioned previously, the $t$-channel amplitude is
very similar to that of sneutrino production; only its final and
intermediate states are exchanged.  The external chargino
wave-function renormalization induces the same $\log M_{\tilde Q}$
correction to the $e$-$\tilde{\nu}_e$-$\tilde\chi^+$ coupling. Also,
the chargino production cross-section is generally large; for the set
of parameters considered, it is near 1 pb for a left-handed electron
beam. Because the chargino cross-section is only proportional to a
single power of $\beta$, the uncertainty in the cross-section due to
the final state mass is effectively reduced by a factor of 3. Notice
$m_{\tilde{\chi}^+_1}$ can be measured through the electron energy
distribution from the $\tilde{\nu}_e$ production and decay. Also, the
chargino production cross-section rises quickly near threshold, which
allows for a determination of its mass by a threshold scan.

In Figs.~6 and 7, we assumed that the decays into the lightest
chargino $\tilde{\chi}^{\pm}_1$ were detectable as $\tilde{\nu}_e$
production signal without background.  This may not be true, depending
on the decay mode of the chargino. For example, if both of the
charginos decay into $q\bar{q}'\tilde\chi_1^0$, it gives an $e^-e^+4j$
signal event with missing energy. Pair production of
$\tilde{e}_L\tilde{e}_L$ could give the same $e^-e^+4j$ signal with
similar kinematics. By SU(2) invariance $\tilde{e}_L$ and
$\tilde{\nu}_e$ are very close in mass, so it is natural to think that
this background is necessarily problematic.  Additionally, at point 3
Ref.~\cite{BMT} estimates that the $\tilde{e}_L$ mass uncertainty is
larger than that of $\tilde{\nu}_e$; this corresponds to a large
uncertainty in the expected $\tilde{e}_L$ production cross-section.
Fortunately, in this case, the cross-section of $\tilde{e}_L$ and its
branching ratio into $\tilde\chi^0_2$ are small, and they are
constrained by other measurements, so they give a negligible
background to the $\tilde\nu_e\tilde\nu_e^*\rightarrow e^-e^+4j$
signal.

We also assumed constant ${\rm BR}(\tilde{\nu}_e\rightarrow
e\tilde\chi ^+_1)$ throughout Figs.~6 and 7, while the branching ratio
is expected to vary with $M_2$, $\mu$, $\tan\beta$,
$m_{\tilde{\ell}}$, and $M_{\tilde Q}$.  Notice, $M_2$, $\mu$, and
$m_{\tilde{\ell}}$ are constrained from mass measurements, while in
Fig.~6 $\tan\beta$ and $M_{\tilde Q}$ are our fit parameters, so the
inclusion of the branching ratio would not introduce additional
uncertainty in the $M_{\tilde Q}$ constraint unless there is a large
correlation.  The chargino branching ratios
($B(\tilde\chi^+\rightarrow jj\tilde\chi^0)$ and
$B(\tilde\chi^+\rightarrow \ell\nu\tilde\chi^0)$) are likewise
determined by the measured sparticle masses and their cross-section
and fit parameters, or they are directly constrained by the study of
chargino and neutralino production at low energy\footnote{When the
chargino cannot decay to an on-shell $W$-boson, the branching ratios
become somewhat sensitive to squark and slepton masses.}.

We did not include the expected luminosity and acceptance
errors. Needless to say, it is extremely important to achieve a low
luminosity error, since we are dealing with a potentially sub-percent
measurement. The effect of the luminosity error can be directly seen
in Fig.~6. In the chargino study of Ref.~\cite{FPMT}, the systematic
error of acceptance is claimed to be significant in the two jet +
lepton final state. There, the large background from $WW$ production
requires a very tight cut, leading to low acceptance and large
systematic error. Here, we consider sneutrino production followed by
decays to charginos. There is very low SM background to the final
state $e^+e^- + (4j {\rm\ or\ } 2j\ell {\rm \ or\ }\ell\ell') + $
missing energy, where the final state electron and positron are hard
and isolated.  Also, the SM background has very different kinematical
structure from the signal. Given the precisely measured chargino,
neutralino and sneutrino masses, it is reasonable to assume the
acceptance error is negligible.

In Ref.~\cite{FPMT}, the possibility to measure $g_{ e\tilde{\nu}_e
\tilde{W}}$ from chargino production has been studied, resulting in a
rather poor constraint $({\cal O}(20\%))$.  In that analysis, the
assumption was made that only chargino production was accessible.
This leads to a rather large uncertainty due to the unknown sneutrino
mass, which could only be determined with a polarized beam through the
differential cross-section of the chargino production itself. That was
a reasonable and conservative assumption at the time, before a
detailed MC simulation of $\tilde{\nu}_e$ production was available.
Including the sneutrino mass measurement will in this case provide a
reasonable constraint on $\bar g$ and hence on the squark mass
\cite{CFPb}; a detailed study will be presented elsewhere \cite{KIYO}.

\section{Conclusions}

Supersymmetry is a beautiful symmetry which relates bosons and
fermions. If we wish to determine whether this symmetry is realized in
nature, the relations imposed between particles and their
superpartners must be confirmed by experiment.  Of course, discovering
a particle with the quantum numbers of a superpartner is the first
very important step in this procedure. An equally important test,
though, is the confirmation of the hard relations imposed by
supersymmetry, for example, the equivalence of the gauge and gaugino
couplings.

It has been argued that a next generation linear collider would be an
excellent tool to verify supersymmetry in this respect.  Production
cross-sections such as $\sigma(e^-e^+\rightarrow
\tilde{\ell}\tilde{\ell}^*)$ and $\sigma(e^-e^+\rightarrow
\tilde{\chi}^-_i\tilde{\chi}^+_i)$ involve the $t$-channel exchange of
gauginos or sleptons, so they depend on gaugino couplings. Several
papers \cite{NFT,FPMT} have shown that it is possible to verify the
equivalence of the gaugino and gauge couplings in a model independent
way, by fitting the data to the cross-sections with arbitrary gaugino
couplings. The sensitivity ranges from about 1\% to 10\%, depending on
the assumptions.

In this paper, we approached this problem from a somewhat different
direction. Because supersymmetry must be badly broken by soft breaking
terms, the tree-level relations of the couplings are also broken, by
radiative corrections. The corrections are logarithmically sensitive
to the splitting of the supersymmetry multiplets. To quantify this, we
have calculated the full one-loop correction due to (s)quark loops of
the slepton production cross-sections. We have explicitly demonstrated
that the full one-loop amplitudes can be factorized into a form which
corresponds to an effective theory in which the heavy squarks are
integrated out. The difference between the effective
lepton-slepton-gaugino couplings
$\bar{g}_{\ell\tilde{\ell}\tilde\chi_i}$ and the effective gauge
couplings $\hat{g}_i^{\rm eff}$ is given by a coupling factor times
$\log M_{\tilde{Q}}/m_{\tilde\ell}$.  The one-loop formulae for the
chargino and neutralino production cross-sections are not given in
this paper, but the effective gaugino couplings receive the same
corrections to logarithmic accuracy.

We gave an explicit example which illustrates that the statistics at
the future linear collider may be enough to constrain the squark mass
scale through the measurement of the slepton production cross-section.
We found, with 1-$\sigma$ significance, $M_{\tilde Q}$ could be
constrained to $1^{+1.2}_{-0.5}$ TeV by the measurement of the
sneutrino production cross-section. This corresponds to a measurement
of the difference between the SU(2) gauge and gaugino couplings
$\delta g_2/g_2 = 0.011\pm0.006$.  We found this constraint in the
$\mu<0$ case where we took into account the errors (based on existing
MC simulation) of the sneutrino and light chargino masses, but assumed
$\tan\beta$ was well constrained by other measurements. In the $\mu>0$
case the slepton cross-sections are much more sensitive to $\tan\beta$
and the heavy (Higgsino dominated) chargino and neutralino masses.  We
need to rely on additional measurements in order to precisely measure
the gaugino couplings in this case.

For either sign of $\mu$, the mass of the sleptons and gauginos must
be measured very precisely in order to successfully constrain the
squark mass scale via production cross-section measurements.  The
effect of the slepton mass uncertainty on the prediction of the
slepton production cross-section is especially significant, as
discussed in Section 4.  In order to determine the ultimate
sensitivity of this procedure, a thorough study of the systematic
uncertainties in the slepton mass measurements is necessary.  The
prediction of the chargino production cross-section is less sensitive
to the uncertainties in the masses.  The limiting factor in the
chargino mode is SM background and statistics.

It is important to note that the constraint on the squark mass scale
can be stronger than the one presented in this paper.  Here, we
estimated the sensitively to squark mass scale by utilizing sneutrino
production followed by its decay into a chargino and an electron.
Depending on the spectrum and center-of-mass energy, there will
typically be many other production processes which involve $t$-channel
exchange of gauginos or sleptons, and all those amplitudes have
$\log M_{\tilde{Q}}$ corrections.

The constraint on the squark mass we have realized here could be
unique in the sense that this information may not be available at the
LHC. Even if the LHC squark production rate is large, the gluino could
be produced in even larger numbers, creating a large irreducible
background to the squark signal. A large gluino background
could make the extraction of the squark mass from kinematical
variables difficult.

On the other hand, if information on the squark masses is obtained at
the LHC, we would have rather accurate predictions for the gaugino
couplings.  In this case, the measurement of the production
cross-sections we considered here would constrain new
supersymmetry-breaking physics with standard model gauge quantum
numbers.  In a sense, the study proposed here is similar in nature to
studies performed at LEP and SLC. The physics of gauge boson two-point
functions has been studied extensively at LEP and SLC, and it has
provided strong constraints on new physics.  Similarly, a future LC
and the LHC might provide precision studies of the gaugino two-point
functions, to realize a supersymmetric version of new precision tests.

\section*{Acknowledgements}
We thank Manuel Drees and Jonathan Feng for useful
discussions.

\section*{Appendix A: One-loop quark/squark corrections}
\setcounter{equation}{0}
\renewcommand{\theequation}{A.\arabic{equation}}

In this section, we show the explicit forms of the quark-squark loop
corrections to the amplitudes of the processes
$e^-(p_1)e^+(p_2)\rightarrow \tilde{\ell}_i(p_3)\tilde{\ell}_j^*(p_4)$
where $\tilde{\ell}_i=(\tilde{e}^-_L, \tilde{e}^-_R, \tilde{\nu}_e)$.
We adopt the \DR~ renormalization scheme, and the \DR~ suffix is implicit
throughout this and the next sections.

The neutralino \DR~ mass matrix $Y_N$ and the chargino mass matrix $Y_C$
are functions of $M_1(Q)$, $M_2(Q)$, $\mu(Q)$, $\tan\beta(Q)$,
$M_W(Q)$, and $M_Z(Q)$ as follows \cite{GH},
\begin{equation}
Y_N(Q)(\tilde{B},\tilde{W^3},\tilde{H}^0_1,\tilde{H}^0_2)\equiv\left(
\begin{array}{cccc}
M_1& 0& -M_Zc_{\beta}s_W & M_Zs_{\beta}s_W\\
0& M_2& M_Z c_{\beta}c_W & -M_Z s_{\beta}c_W\\
-M_Z c_{\beta}s_W & M_Z c_{\beta}c_W & 0 & -\mu\\
M_Z s_{\beta}s_W & -M_Z s_{\beta}c_W & -\mu & 0         
\end{array}
\right)\ .                
\end{equation}
Here $c_{\beta}\equiv\cos\beta$, $s_{\beta}\equiv\sin\beta$,
$c_W\equiv M_W/M_Z$ and $s_W\equiv\sqrt{1-c_W^2}$. The $Q$ dependence
of the right hand side is implicit. A unitary matrix $N$ is defined so
that $M^D_N=(m_{\tilde{\chi}^0_i}\delta_{ij})= N^* Y_N N^{\dagger}$.
The chargino \DR~ mass matrix is
\begin{equation}
Y_C(Q)(\tilde{W},\tilde{H})\equiv\left(
\begin{array}{cc}
M_2& \sqrt{2}M_W s_{\beta}\\
\sqrt{2}M_W c_{\beta} & \mu 
\end{array}
\right)\ ,
\end{equation}
and the chargino mass matrix can be diagonalized by two unitary
matrices $U$ and $V$ as $M^D_C\equiv U^* Y_C V^{\dagger}$.  We order
the charginos and neutralinos according to their masses, so,
for example, $\tilde\chi^0_1$ is the lightest neutralino.

We first show the $t$-channel amplitudes.  They are expressed in terms
of two-point functions $iK_{0ij}$, $iK_{-ij}$ of neutralinos
$\tilde{\chi}^0_i\tilde{\chi}^0_j$ and charginos
$\tilde{\chi}^+_i\tilde{\chi}^-_j$.  $K_{ij}$ are decomposed as
\begin{equation}
K_{ij}=\Sigma^L_{ij}(p^2)\psla P_L+\Sigma^R_{ij}(p^2)\psla P_R
+\Sigma^M_{ij}(p^2)P_L+\Sigma^{M*}_{ij}(p^2)P_R\ , \label{eqa1}
\end{equation}
where the momentum $p$ flows from $j$ to $i$.  $P_{R,L}$
are the chiral projectors, $(1\pm\gamma_5)/2$.

In the following the argument of the $\Sigma$ functions is $p^2$. The
corrected $t$-channel amplitude $i{\cal M}_{LL}^{(t)}$ for
$\tilde{e}^-_L\tilde{e}^+_L$ is then expressed as
\begin{eqnarray}
i{\cal M}_{LL}^{(t)} &=& -i \bar{v}(p_2) \psla P_L u(p_1) 
a^{0*}_{\tilde{e}Li}a^0_{\tilde{e}Lj}
\left[ \frac{\delta_{ij}}{p^2-m_{\tilde{\chi}^0_i}^2} \right. 
\nonumber \\
&&\left. - {1\over(p^2-m_{\tilde{\chi}^0_i}^2)
(p^2-m_{\tilde{\chi}^0_j}^2)}
\biggl(m_{\tilde{\chi}^0_i} m_{\tilde{\chi}^0_j} 
\Sigma^L_{0ij} +p^2 \Sigma^R_{0ij}+m_{\tilde{\chi}^0_j} \Sigma^M_{0ij}
+m_{\tilde{\chi}^0_i} \Sigma^{M*}_{0ij}\biggr)\right]\ , \label{eqa2}
\end{eqnarray}
where $p=p_1-p_3=p_4-p_2$. Similarly, other $t$-channel amplitudes are 
\begin{eqnarray}
i{\cal M}_{RR}^{(t)} &=& -i \bar{v}(p_2) \psla P_R u(p_1) 
b^{0*}_{\tilde{e}Ri}b^0_{\tilde{e}Rj}
\left[ \frac{\delta_{ij}}{p^2-m_{\tilde{\chi}^0_i}^2} \right.
\nonumber \\
&& \left. -\frac{1}{(p^2-m_{\tilde{\chi}^0_i}^2)(p^2-m_{\tilde{\chi}^0_j}^2)}
(p^2 \Sigma^L_{0ij}+ m_{\tilde{\chi}^0_i} m_{\tilde{\chi}^0_j} 
\Sigma_{0ij}^R +m_{\tilde{\chi}^0_i} \Sigma^M_{0ij}
+m_{\tilde{\chi}^0_j} \Sigma^{M*}_{0ij})\right]\ , \label{eqa3} \\
i{\cal M}_{LR}^{(t)} &=& -i \bar{v}(p_2) P_L u(p_1) 
b^{0*}_{\tilde{e}Ri}a^0_{\tilde{e}Lj}
\left[ \frac{m_{\tilde{\chi}^0_i}\delta_{ij}}{p^2-m_{\tilde{\chi}^0_i}^2} 
\right. \nonumber \\
&& \left. -\frac{1}{(p^2-m_{\tilde{\chi}^0_i}^2)(p^2-m_{\tilde{\chi}^0_j}^2)}
(p^2(m_{\tilde{\chi}^0_j} \Sigma_{0ij}^L +m_{\tilde{\chi}^0_i} 
\Sigma^R_{0ij})+m_{\tilde{\chi}^0_i}m_{\tilde{\chi}^0_j} \Sigma^M_{0ij}
+p^2 \Sigma^{M*}_{0ij})\right]\ , \label{eqa4} \\ 
i{\cal M}_{RL}^{(t)} &=& -i \bar{v}(p_2) P_R u(p_1) 
a^{0*}_{\tilde{e}Li}b^0_{\tilde{e}Rj}
\left[ \frac{m_{\tilde{\chi}^0_i}\delta_{ij}}{p^2-m_{\tilde{\chi}^0_i}^2} 
\right. \nonumber \\
&& \left. -\frac{1}{(p^2-m_{\tilde{\chi}^0_i}^2)(p^2-m_{\tilde{\chi}^0_j}^2)}
(p^2(m_{\tilde{\chi}^0_i} \Sigma_{0ij}^L +m_{\tilde{\chi}^0_j} 
\Sigma^R_{0ij})
+p^2 \Sigma^M_{0ij}+m_{\tilde{\chi}^0_i}m_{\tilde{\chi}^0_j} 
\Sigma^{M*}_{0ij})\right]\ , \label{eqa5} \\ 
i{\cal M}_{\nu\nu}^{(t)} &=& -i \bar{v}(p_2) \psla P_L u(p_1) 
a^{-*}_{\tilde{\nu}Li}a^-_{\tilde{\nu}Lj}
\left[ \frac{\delta_{ij}}{p^2-m_{\tilde{\chi}^+_i}^2} \right. 
\nonumber \\
&& \left. - \frac{1}{(p^2-m_{\tilde{\chi}^+_i}^2)
(p^2-m_{\tilde{\chi}^+_j}^2)}
(m_{\tilde{\chi}^+_i} m_{\tilde{\chi}^+_j} 
\Sigma_{-ij}^L +p^2 \Sigma^R_{-ij}+m_{\tilde{\chi}^+_j} \Sigma^M_{-ij}
+m_{\tilde{\chi}^+_i} \Sigma^{M*}_{-ij})\right]\ , \label{eqa6} 
\end{eqnarray}
Here $(a,b)$ denote the couplings of 
neutralinos and charginos to fermions and sfermions, 
\begin{eqnarray}
{\cal L}_{\rm int}&=&-\tilde{f}_i^*\bar{\tilde{\chi}}^0_j
(a^0_{\tilde{f}ij}P_L+b^0_{\tilde{f}ij}P_R)f
+({\rm h.c.}) \nonumber \\
&&-\tilde{f_1}_i^*\overline{\tilde{\chi}_j^-}
(a^-_{\tilde{f_1}ij}P_L+b^-_{\tilde{f_1}ij}P_R)f_2
+({\rm h.c.}) \nonumber \\
&&-\tilde{f_2}_i^*\overline{\tilde{\chi}_j^+}
(a^+_{\tilde{f_2}ij}P_L+b^+_{\tilde{f_2}ij}P_R)f_1
+({\rm h.c.})\ , \label{eqa7} 
\end{eqnarray}
where $f=(q,l)$ and $(f_1,f_2)$ are SU(2) doublets. 
Explicit forms of $(a,b)$ for the gauge eigenstates of $\tilde{f}$ 
are written in terms of mixing matrices $(N,U,V)$ and Yukawa 
couplings $y_f$ of fermions $f$ as 
\begin{eqnarray}
&& a^0_{\tilde{f}Li}=\sqrt{2}(g_2I_{3f_L}N^*_{i2}
+g_YY_{f_L}N^*_{i1})\ , \nonumber\\
&& a^0_{\tilde{f_1}Ri}=y_{f_1}N^*_{i4}\ ,\;
a^0_{\tilde{f_2}Ri}=y_{f_2}N^*_{i3}, \nonumber \\
&& b^0_{\tilde{f}Ri}=-\sqrt{2}g_YY_{f_R}N_{i1}, \nonumber\\
&& b^0_{\tilde{f_1}Li}=y_{f_1}N_{i4}\ ,\;
b^0_{\tilde{f_2}Li}=y_{f_2}N_{i3}\ , \nonumber \\
&& a^-_{\tilde{f_1}Li}=g_2V^*_{i1}\ ,\; 
a^+_{\tilde{f_2}Li}=g_2U^*_{i1}\ , \nonumber\\
&& a^-_{\tilde{f_1}Ri}=-y_{f_1}V^*_{i2}\ ,\; 
a^+_{\tilde{f_2}Li}=-y_{f_2}U^*_{i2}\ , \nonumber\\
&& b^-_{\tilde{f_1}Li}=-y_{f_2}U_{i2}\ ,\; 
b^+_{\tilde{f_2}Li}=-y_{f_1}V_{i2}\ ,\nonumber \\ 
&& b^{\mp}_{\tilde{f}Ri}=0\ .
\label{eqa8}
\end{eqnarray}

We next show the explicit forms of the two-point functions 
$\Sigma_{0ij}$ and $\Sigma_{-ij}$ from quark-squark loops. 
For neutralinos, we have 
\begin{eqnarray}
\Sigma^L_{0ij}(p^2)&=& 
\frac{N_c}{16\pi^2}(a^{0*}_{\tilde{q}ki}a^0_{\tilde{q}kj}+
b^0_{\tilde{q}ki}b^{0*}_{\tilde{q}kj})
B_1(p^2, m_q, m_{\tilde{q}_k})\ , \nonumber\\
\Sigma^R_{0ij}(p^2)&=&\Sigma^L_{0ji}(p^2)\ , \nonumber\\
\Sigma^M_{0ij}(p^2)&=& 
\frac{N_c}{16\pi^2}(b^{0*}_{\tilde{q}ki}a^0_{\tilde{q}kj}+
a^0_{\tilde{q}ki}b^{0*}_{\tilde{q}kj})m_q
B_0(p^2, m_q, m_{\tilde{q}_k})\ . \label{eqa9}
\end{eqnarray}
For charginos we have 
\begin{eqnarray}
\Sigma^L_{-ij}(p^2)&=& 
\frac{N_c}{16\pi^2}(
a^{+*}_{\tilde{d}ki}a^+_{\tilde{d}kj}B_1(p^2, m_u, m_{\tilde{d}_k})+
b^-_{\tilde{u}ki}b^{-*}_{\tilde{u}kj}B_1(p^2, m_d, m_{\tilde{u}_k}))\ ,
\nonumber\\
\Sigma^R_{-ij}(p^2)&=& 
\frac{N_c}{16\pi^2}(
b^{+*}_{\tilde{d}ki}b^+_{\tilde{d}kj}B_1(p^2, m_u, m_{\tilde{d}_k})+
a^-_{\tilde{u}ki}a^{-*}_{\tilde{u}kj}B_1(p^2, m_d, m_{\tilde{u}_k}))\ ,
\nonumber\\
\Sigma^M_{-ij}(p^2)&=& 
\frac{N_c}{16\pi^2}(
b^{+*}_{\tilde{d}ki}a^+_{\tilde{d}kj}m_uB_0(p^2, m_u, m_{\tilde{d}_k})+
a^-_{\tilde{u}ki}b^{-*}_{\tilde{u}kj}m_dB_0(p^2, m_d, m_{\tilde{u}_k}))\ . 
\label{eqa10}
\end{eqnarray}
Here $B_{0,1}$ are 't Hooft-Veltman functions in the convention of 
\cite{P}.

Finally, we show the corrected $s$-channel amplitudes 
$i{\cal M}_{\tilde{\ell}\tilde{\ell}^*}^{(s)}$ for 
$e^-(p_1)e^+(p_2)\rightarrow \tilde{\ell}_i(p_3)\tilde{\ell}_i^*(p_4)$. 
In terms of the transverse part of the \DR~ gauge-boson self-energies,
$\Pi^T_{\gamma\gamma, \gamma Z, ZZ}(s)$, they are expressed as 
\begin{eqnarray}
i{\cal M}_{\tilde{\ell}\tilde{\ell}^*}^{(s)}&=&
+\ 2ie^2Q_eQ_{\tilde{\ell}}\left(1- \frac{\Pi^T_{\gamma\gamma}(s)}{s}\right)
\bar{v}(p_2)\psla u(p_1) \nonumber\\
&&+2ieg_Z\frac{\Pi^T_{\gamma Z}(s)}{s(s-M_Z^2)}
\bar{v}(p_2)\psla\biggl[Q_e(I_{3\tilde{\ell}}-s_W^2Q_{\tilde{\ell}})+  
(I_{3e}P_L-s_W^2Q_e)Q_{\tilde{\ell}}\biggr]u(p_1) \nonumber\\
&& 
+\ 2ig_Z^2(I_{3\tilde{\ell}}-s_W^2Q_{\tilde{\ell}})
\frac{1}{s-M_Z^2}
\left( 1-\frac{\Pi^T_{ZZ}(s)-\Pi^T_{ZZ}(M_Z^2)}{s-M_Z^2} \right) 
\nonumber \\
&&\times \bar{v}(p_2) \psla (I_{3e}P_L-s_W^2Q_e)u(p_1),      \label{eqa11}
\end{eqnarray}
where $s=(p_1+p_2)^2$, $g_Z=g_2/c_W$, and $M_Z$ is the $Z$-boson pole
mass.

\section*{Appendix B: Effective couplings for gauginos}
\setcounter{equation}{0}
\renewcommand{\theequation}{B.\arabic{equation}}

In this appendix, we show how the full one-loop $t$-channel amplitudes
factorize into a form corresponding to an effective theory. The
effective theory amplitude depends on effective couplings, masses, and
mixing matrices. We explicitly show what approximations are necessary
for this, and justify them. As a by-product, we show the cancellation
of UV divergences in the one-loop amplitudes. Here, for brevity, only
the case of $\tilde{e}^-_R\tilde{e}^+_R$ production is discussed. The
generalization to other channels is straightforward.

We first prove that the loop corrected $t$-channel amplitude
(Eq.~(\ref{eqa3})) can be written in terms of the mass matrix of
neutralinos $Y\equiv Y_N$ in the gauge eigenbasis, instead of the
mixing matrix $N$ and $m_{\tilde{\chi}^0_i}$.  By substituting
Eq.~(\ref{eqa8}), Eq.~(\ref{eqa3}) takes the form (the $\Sigma_{ij}
\equiv\Sigma_{0_{ij}}$'s are functions of $p^2$)
\begin{eqnarray}
i{\cal M}_{RR}^{(t)} &=& g_Y^2 N^*_{i1}
\left[ \frac{\delta_{ij}}{p^2-m_i^2} \right. 
\nonumber \\
&&\left. - \frac{1}{(p^2-m_i^2)(p^2-m_j^2)}
(p^2 \Sigma^L_{ij}+ m_i m_j \Sigma_{ij}^R +m_i \Sigma^M_{ij}
+m_j \Sigma^{M*}_{ij})\right] N_{j1}X\ . \label{eqb1}
\end{eqnarray}
Here we use abbreviations $m_i=m_{\tilde{\chi}^0_i}$ and 
$X=-2iY_{e_R}^2\bar{v}(p_2) \psla P_R u(p_1)$.

At tree-level, $Y$, diagonalized masses 
$M_D={\rm diag}(m_i)$, and $N$ are related by
\begin{equation}
M_D=N^*YN^{\dagger}=NY^{\dagger}N^T\ . \label{eqb2}
\end{equation}
We can transform the tree-level part of Eq.~(\ref{eqb1}) as 
\begin{eqnarray}
N_{i\alpha}^*\frac{1}{p^2-m_i^2}N_{i\beta} &=& 
\left[ N^{\dagger}\frac{1}{p^2-M_D^2}N \right]_{\alpha\beta} 
\nonumber \\
&=& \left[ N^{\dagger}\frac{1}
{p^2-NY^{\dagger}YN^{\dagger}}N \right]_{\alpha\beta} \nonumber \\
&=& \left[ \frac{1}{p^2-Y^{\dagger}Y} \right]_{\alpha\beta}\ ,
\label{eqb3}
\end{eqnarray}
where $\alpha=\beta=1$ in this case. The loop corrected part 
can be written by noting that the two-point functions $\Sigma$ 
are given in terms of the gauge eigenbasis functions $\tilde\Sigma$, as
\begin{eqnarray}
\Sigma^L &=& N\tilde{\Sigma}^LN^{\dagger}, \nonumber\\
\Sigma^R &=& N^*\tilde{\Sigma}^RN^T, \nonumber\\
\Sigma^M &=& N^*\tilde{\Sigma}^MN^{\dagger}. \label{eqb4}
\end{eqnarray}

The total $t$-channel amplitude is therefore expressed as 
\begin{eqnarray}
i{\cal M}_{RR}^{(t)} &=& g_Y^2 
\left[ (p^2-Y^{\dagger}Y)^{-1}- 
(p^2-Y^{\dagger}Y)^{-1} \left\{ 
(p^2 \tilde{\Sigma}^L(p^2) \right. \right. 
\nonumber \\
&&\left. \left. +Y^{\dagger}\tilde{\Sigma}^R(p^2)Y 
+Y^{\dagger}\tilde{\Sigma}^M(p^2)
+\tilde{\Sigma}^{M*}(p^2)Y \right\} (p^2-Y^{\dagger}Y)^{-1}
\right]_{11} X\ , \label{eqb5}
\end{eqnarray}

Next, we absorb the main loop contributions into an effective
neutralino mixing matrix and effective couplings.  We define the
effective neutralino mass matrix as
\begin{equation}
\overline{Y}(p^2)=Y-\tilde\Sigma^M(p^2)
-\frac{1}{2}Y\tilde\Sigma^L(p^2) -\frac{1}{2}
\tilde\Sigma^R(p^2)Y\ . \label{eqb6}
\end{equation}
The one-loop pole
masses $m_i({\rm pole})$ are obtained from $\overline{Y}$ as
\begin{eqnarray} m_i({\rm pole})&=& {\rm
Re}(N^*\overline{Y}(m_i^2)N^{\dagger})_{ii} \nonumber \\ &=&m_i
-\frac{1}{2}[m_i(\Sigma^L+\Sigma^R)_{ii}
+(\Sigma^M+\Sigma^{M*})_{ii}](m_i^2)\ , \end{eqnarray}
The amplitude Eq.~(\ref{eqb5}) becomes, up to ${\cal O}(g_Y^2)$, 
\begin{eqnarray}
i{\cal M}_{RR}^{(t)} &=& g_Y^2 X 
\left[ (p^2-\overline{Y}^{\dagger}\overline{Y}(p^2))^{-1} 
\right. \nonumber\\
&& \left. 
+\frac{1}{2}(p^2-Y^{\dagger}Y)^{-1}\tilde\Sigma^L
+\frac{1}{2}\tilde\Sigma^L(p^2-Y^{\dagger}Y)^{-1} \right] _{11}\ .
\label{eqb7}
\end{eqnarray}
The remaining $\tilde\Sigma^L_{ij}$ have an important
property. Their off-diagonal terms are generated by the breaking of
SU(2)$\times$U(1) gauge symmetry in the quark-squark sector, and are
UV finite. For sufficiently heavy squarks the contributions of the
factors $\tilde\Sigma^L_{1i,i1}(i\ne 1)$ are insignificant and can
be dropped.  Then, by introducing the effective gaugino coupling
\begin{equation}
\bar{g}_{e\tilde{e}_R\tilde{B}}(p^2)
\equiv g_Y\left( 1-\frac{1}{2}\tilde\Sigma^L_{11}(p^2)\right)\ ,
\label{eqb8}
\end{equation}
the $t$-channel amplitude is expressed in a very simple form 
\begin{eqnarray}
i{\cal M}_{RR}^{(t)} &\sim & \overline{g}^2_{e\tilde{e}_R\tilde{B}}(p^2) 
[(p^2-\overline{Y}^{\dagger}\overline{Y}(p^2))^{-1}]_{11}X
\nonumber\\
&& = \left( \frac{\bar{g}^2_{e\tilde{e}_R\tilde{B}}
\overline{N}_{i1}^*\overline{N}_{i1}(p^2) }
{p^2-\overline{m}_i^2(p^2)} \right)X\ , \label{eqb9}
\end{eqnarray}
where $\overline{m}_i(p^2)$ and $\overline{N}(p^2)$ are the effective
masses and the effective mixing matrix, respectively, obtained by
diagonalizing $\overline{Y}(p^2)$. Eq.~(\ref{eqb9}) takes the same
form as the tree-level amplitude Eq.~(\ref{eqb1}) with the replacement
of the neutralino masses and couplings with the effective ones.  It is
now evident that the corrected amplitude Eq.~(\ref{eqb1}) is UV
finite, since both $\overline{Y}(p^2)$ and
$\bar{g}_{e\tilde{e}_R\tilde{B}}(p^2)$ are finite.

The $t$-channel amplitudes for other processes are similarly expressed
in terms of effective mass matrices and couplings of neutralinos and
charginos.  Below we list the forms of other effective
lepton-slepton-gaugino couplings which are relevant to our processes:
\begin{eqnarray}
\bar{g}_{e\tilde e_L\tilde B}(p^2)
&=& g_Y\left(1-\frac{1}{2}{\tilde\Sigma^R_{0_{11}}}(p^2)\right)\ , 
\nonumber\\
\bar{g}_{e\tilde{e}_L\tilde{W}}(p^2)
&=& g_2\left( 1-\frac{1}{2}{\tilde{\Sigma}^R_{0_{22}}}(p^2)\right)\ , 
\nonumber\\
\bar{g}_{e\tilde{\nu}_e\tilde{W}}(p^2)
&=& g_2\left( 1-\frac{1}{2}{\tilde{\Sigma}^R_{-_{11}}}(p^2)\right)\ . 
\label{eqb10}
\end{eqnarray}
Note that the effective couplings are independent of (s)fermion
flavor, and hence the $(\log M_{\tilde Q}+{\rm const})$ corrections
are universal, in analogy with the oblique corrections of the gauge
bosons \cite{CFPa,RKS}.

\newpage

\begin{figure}[htb]
\includegraphics{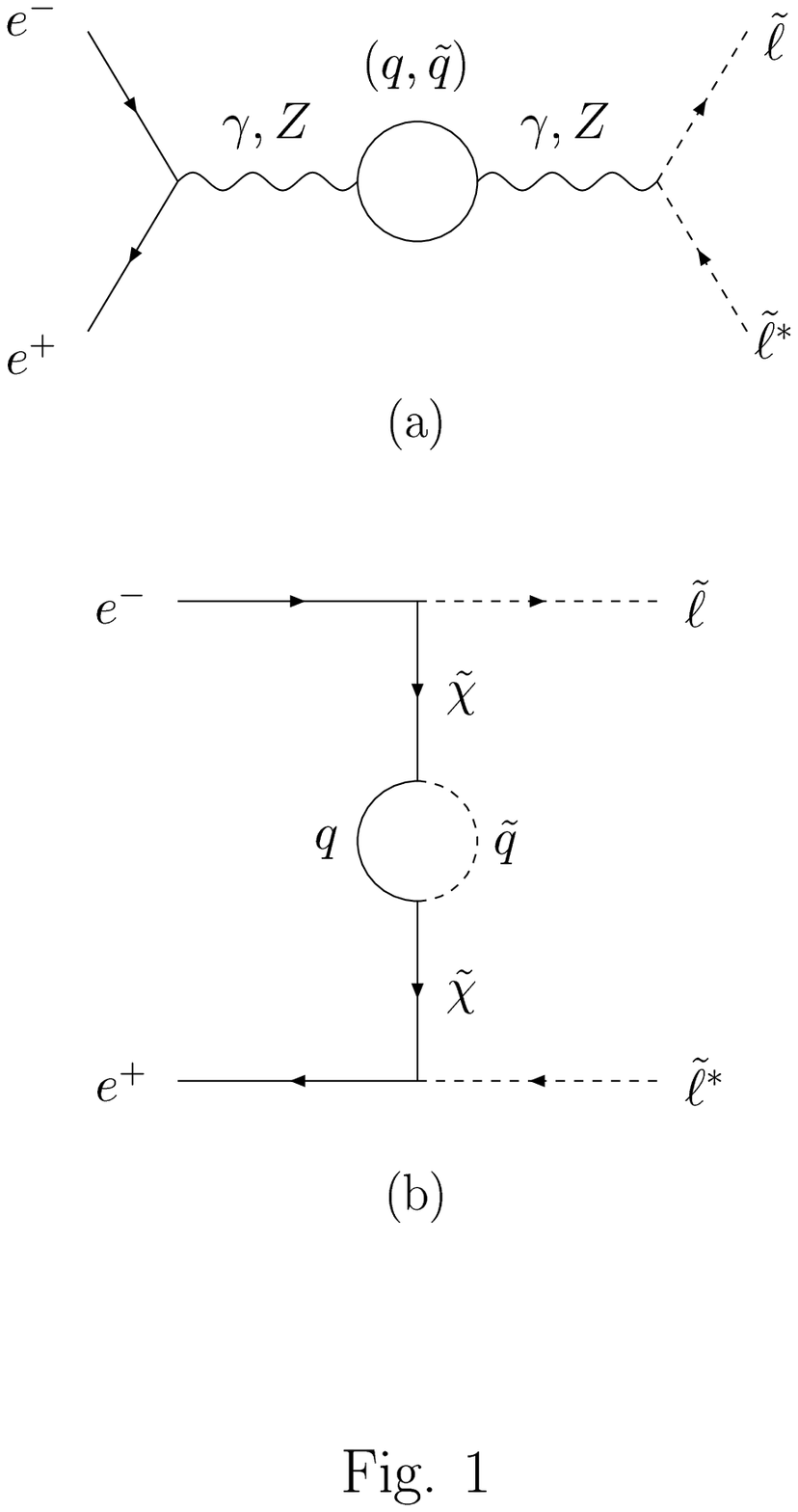}
\caption{Feynman graphs of the one-loop quarks-squark corrections to 
the processes $e^-e^+\rightarrow\tilde{\ell}\tilde{\ell}^*$, 
for (a) $s$-channel and (b) $t$-channel amplitudes. }
\end{figure}

\newpage

\begin{figure}[htb]
\includegraphics{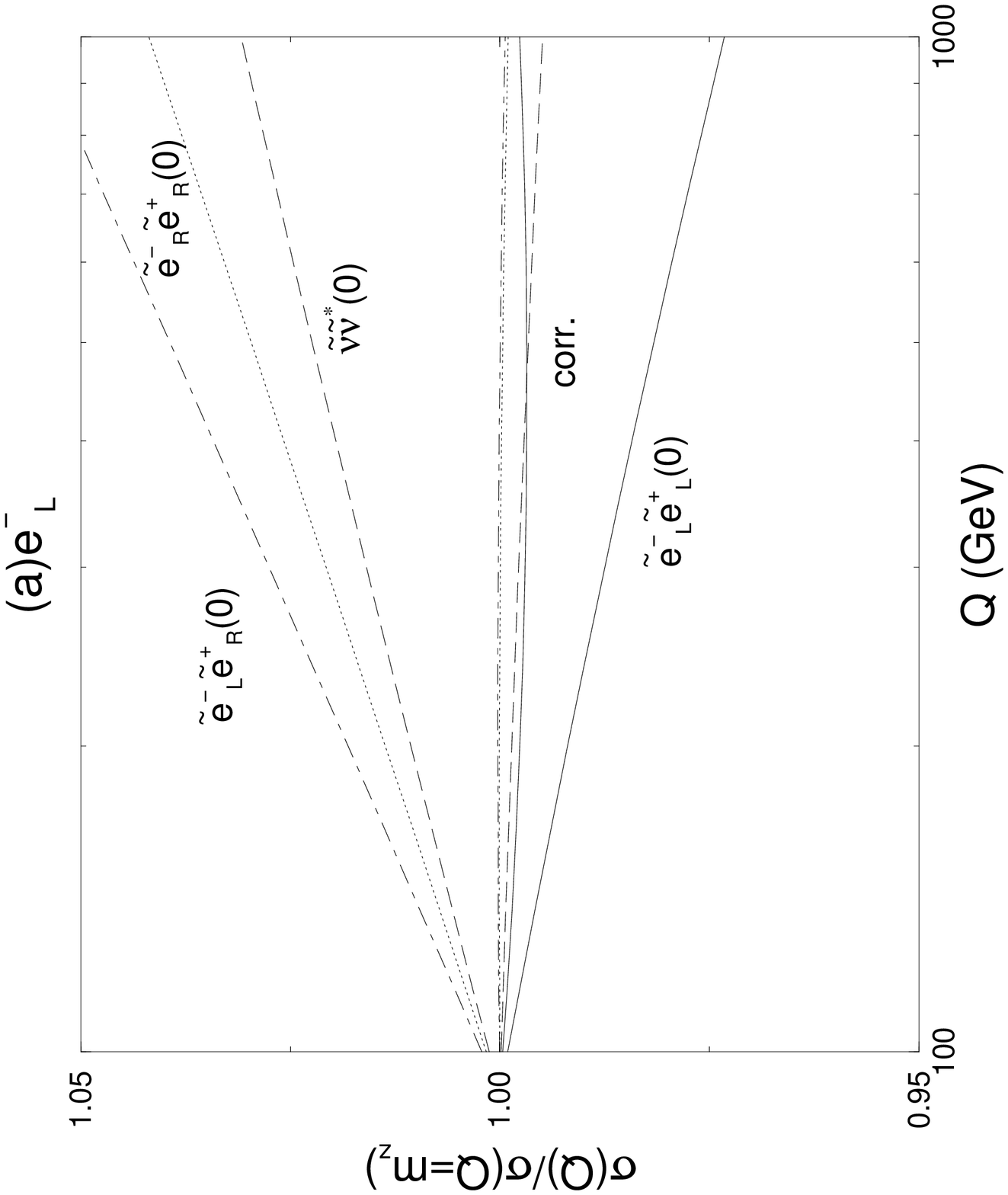}
\caption{ The dependence of the lowest-order (0) and one-loop (corr)
cross-sections of all possible slepton production channels on the
\protect\DR~ renormalization scale $Q$. The ratios
$\sigma(Q)/\sigma(Q=M_Z)$ are shown for each channel.  Parameters are
chosen as $M_1(M_1)=100$~GeV, $M_2(M_2)=200$~GeV, $\mu(|\mu|)=-300$~GeV,
$\tan\beta(M_Z)=4$, $m_{\tilde{\ell}}=200$~GeV,
$M_{\tilde{Q}}=1000$~GeV, and $\protect\sqrt{s}=500$~GeV.  }
\includegraphics{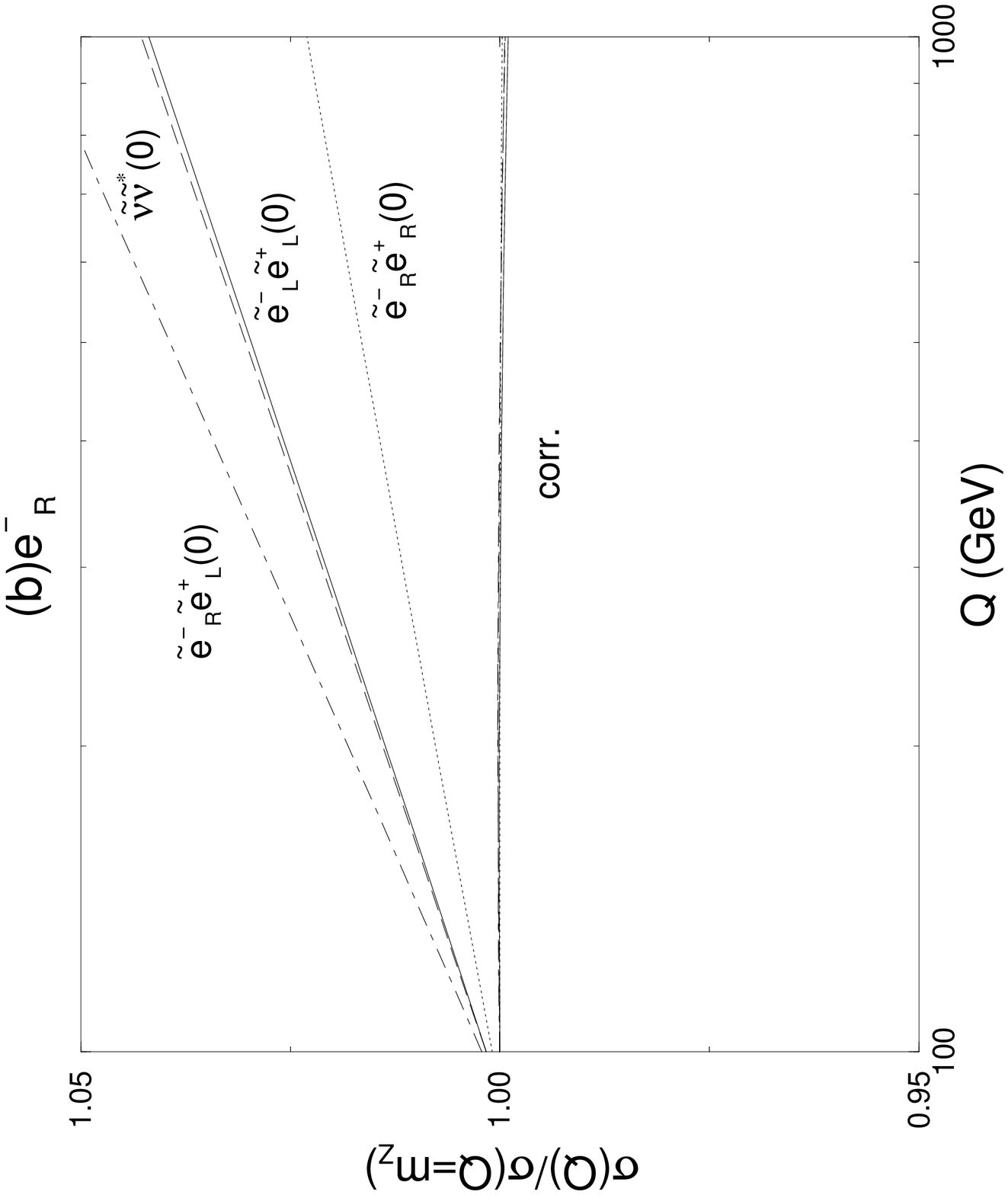}
\end{figure}

\newpage

\begin{figure}[htb]
\includegraphics{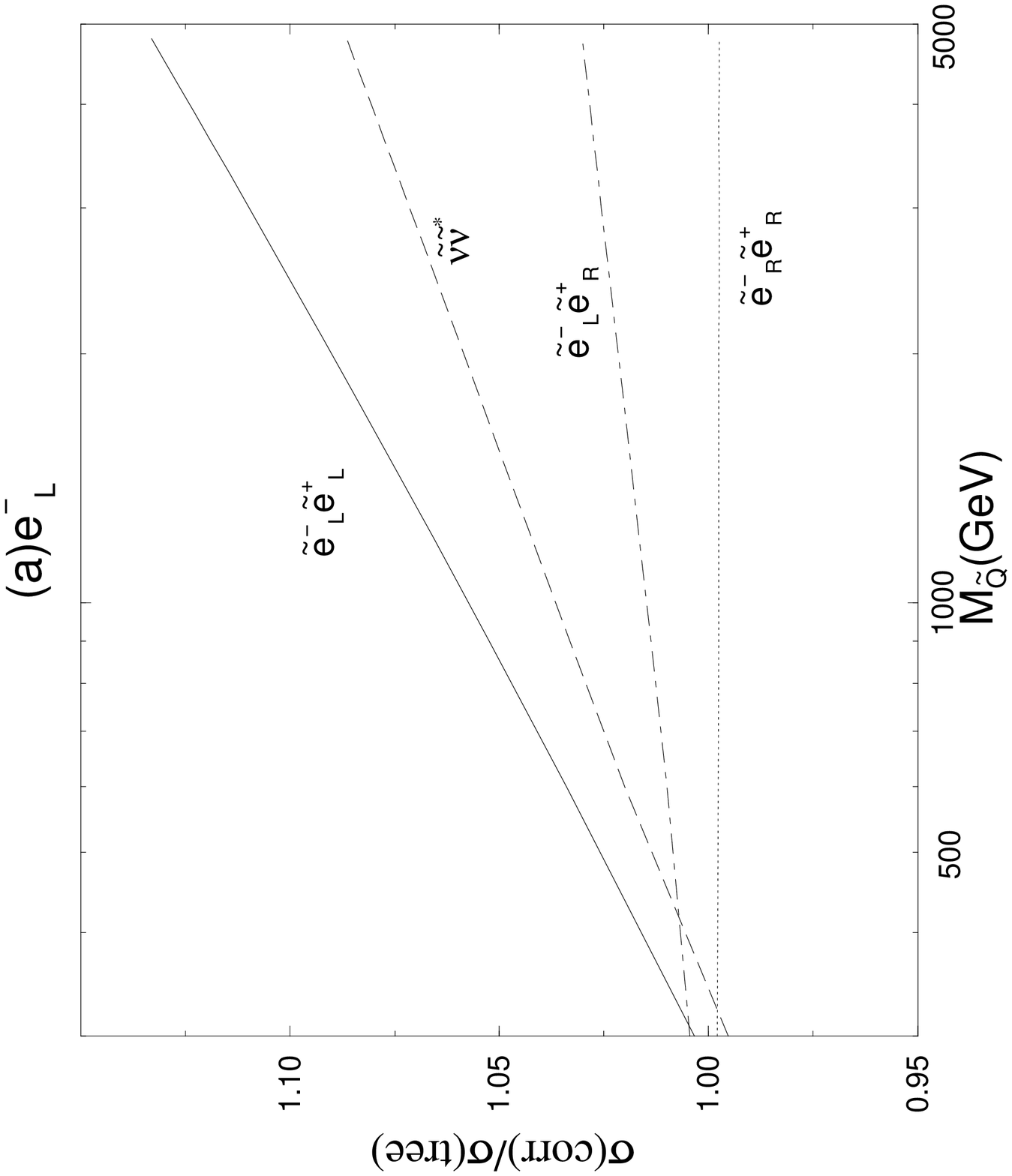}
\caption{ 
The $M_{\tilde{Q}}$ dependence of the slepton production
cross-sections. Input parameters are 
$m_{\tilde{\chi}^0_1}=100$~GeV, $m_{\tilde{\chi}^+_1}=200$~GeV, 
$m_{\tilde{\chi}^0_3}=300$~GeV, $\tan\beta(M_Z)=4$, 
$m_{\tilde{\ell}}=200$~GeV, $A=0$, $\mu<0$, and $\protect
\sqrt{s}=500$~GeV. 
The corrected cross-sections are normalized by the tree-level
cross-sections, which are defined in the text.}
\includegraphics{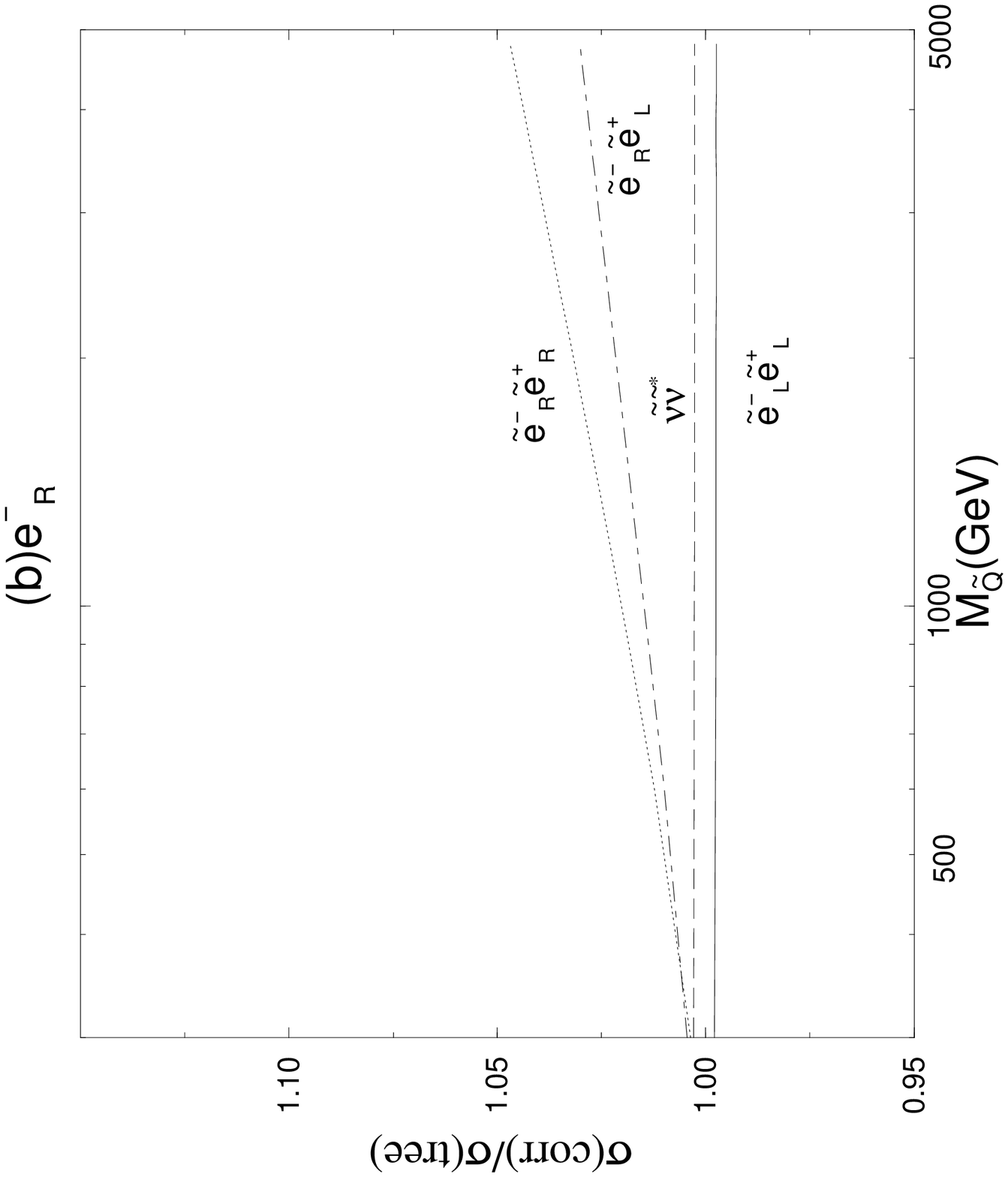}
\end{figure}

\newpage

\begin{figure}[htb]
\includegraphics{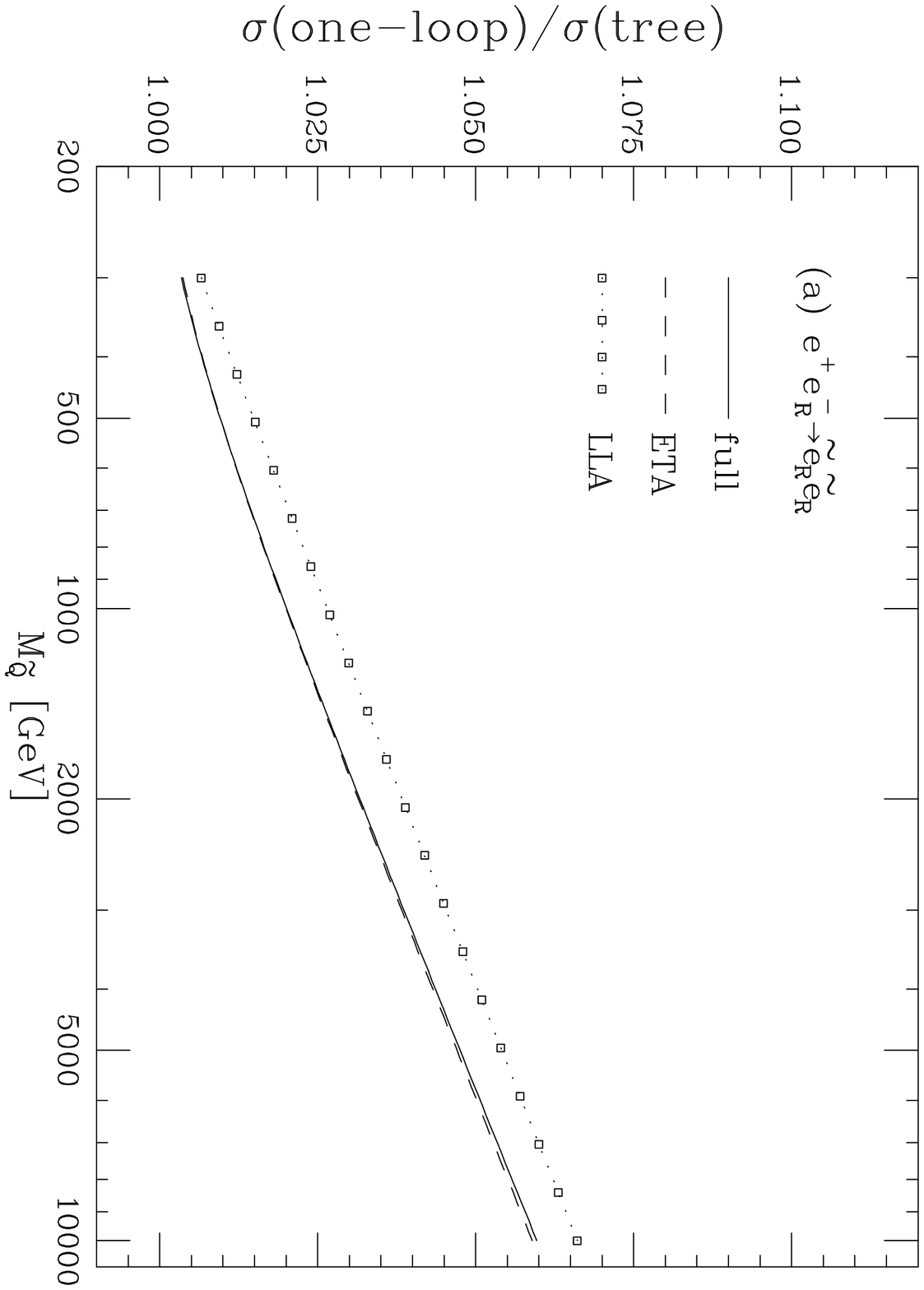}
\caption{The full corrected cross-sections (solid lines) and two
approximations, the effective theory approximation (ETA, dashed
lines), and leading log approximation (LLA, dotted with squares), for
two channels $e^-_Le^+\rightarrow$ $\tilde{\nu}_e\tilde{\nu}_e^*$ and
$e^-_Re^+\rightarrow$ $\tilde{e}^-_R\tilde{e}^+_R$, vs.
$M_{\tilde{Q}}$. The cross-sections are normalized by the tree-level
ones, as in Fig.~3.  Input parameters are the same as in Fig.~3. }
\includegraphics{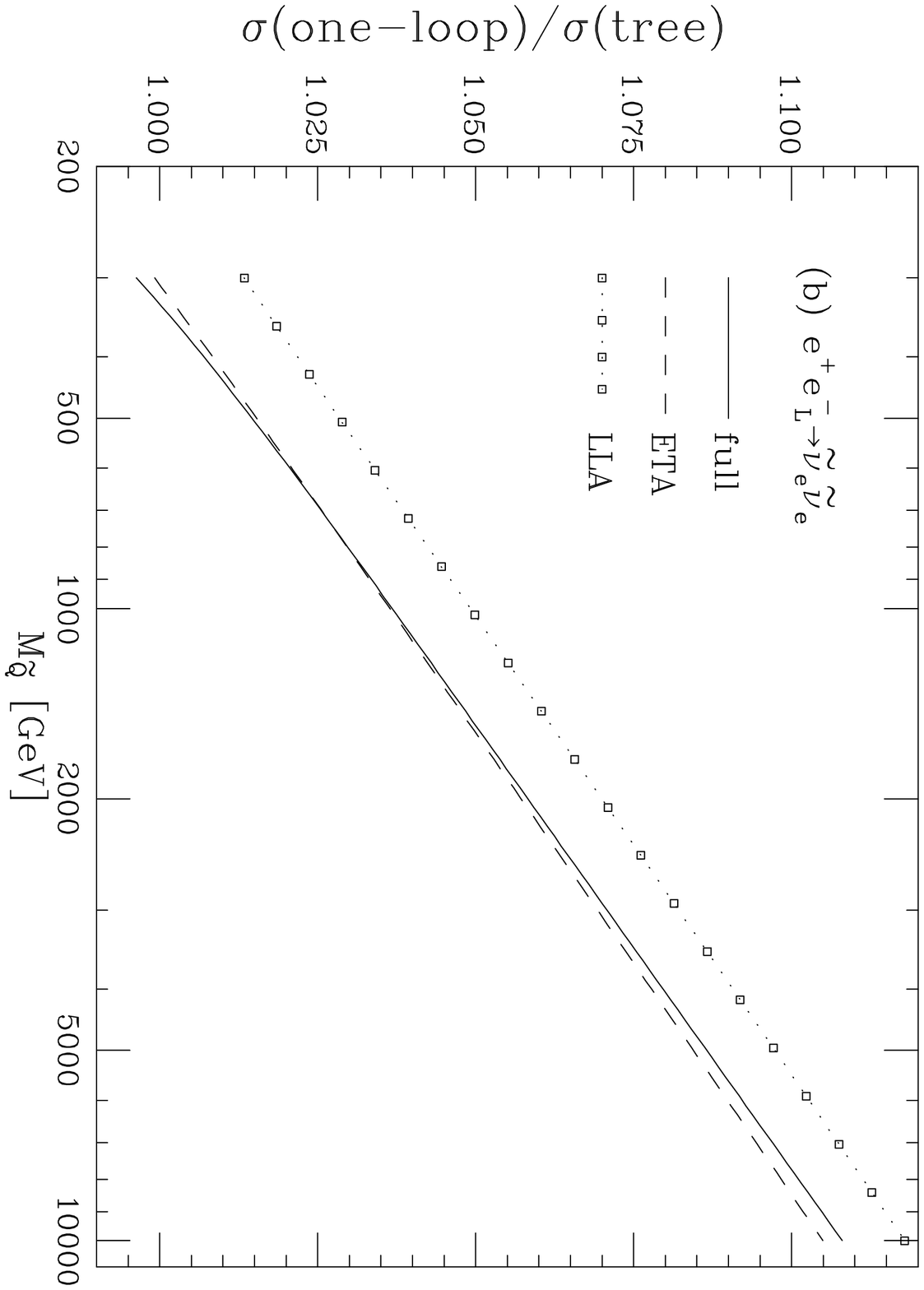}
\end{figure}

\newpage

\begin{figure}[htb]
\includegraphics{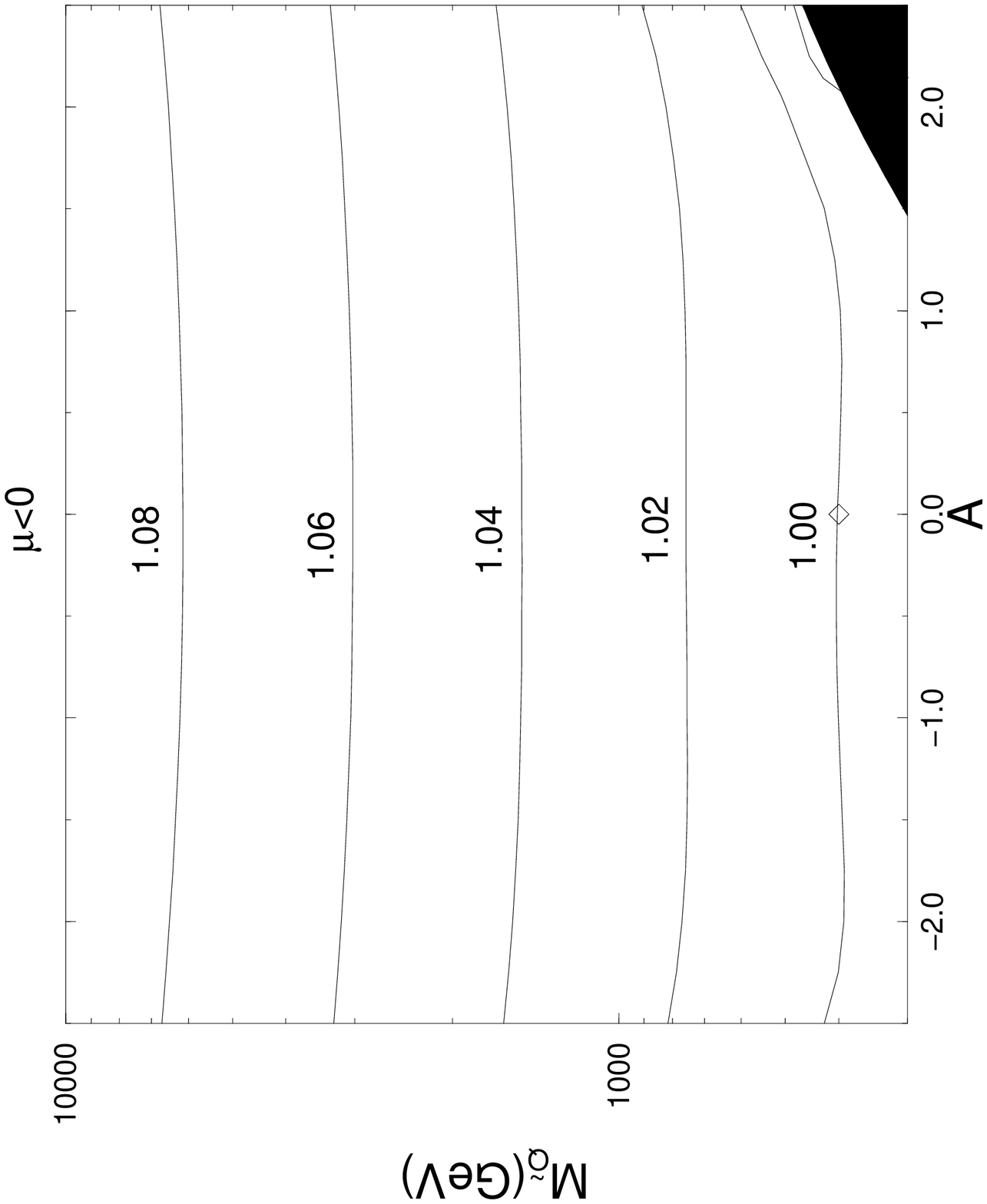}
\caption{ The cross-section
$\sigma(e^-_Le^+\rightarrow\tilde{\nu}_e\tilde{\nu}_e^*)$ as a
function of $(M_{\tilde{Q}},A)$, normalized by the value at
$M_{\tilde{Q}}=400$~GeV and $A=0$. Input parameters are the same as in
Fig.~3, except that $\tan\beta(M_Z)=1.5$ and $\mu$ is either (a)
negative or (b) positive. In the black regions a squark is lighter
than the $Z$-boson.}
\includegraphics{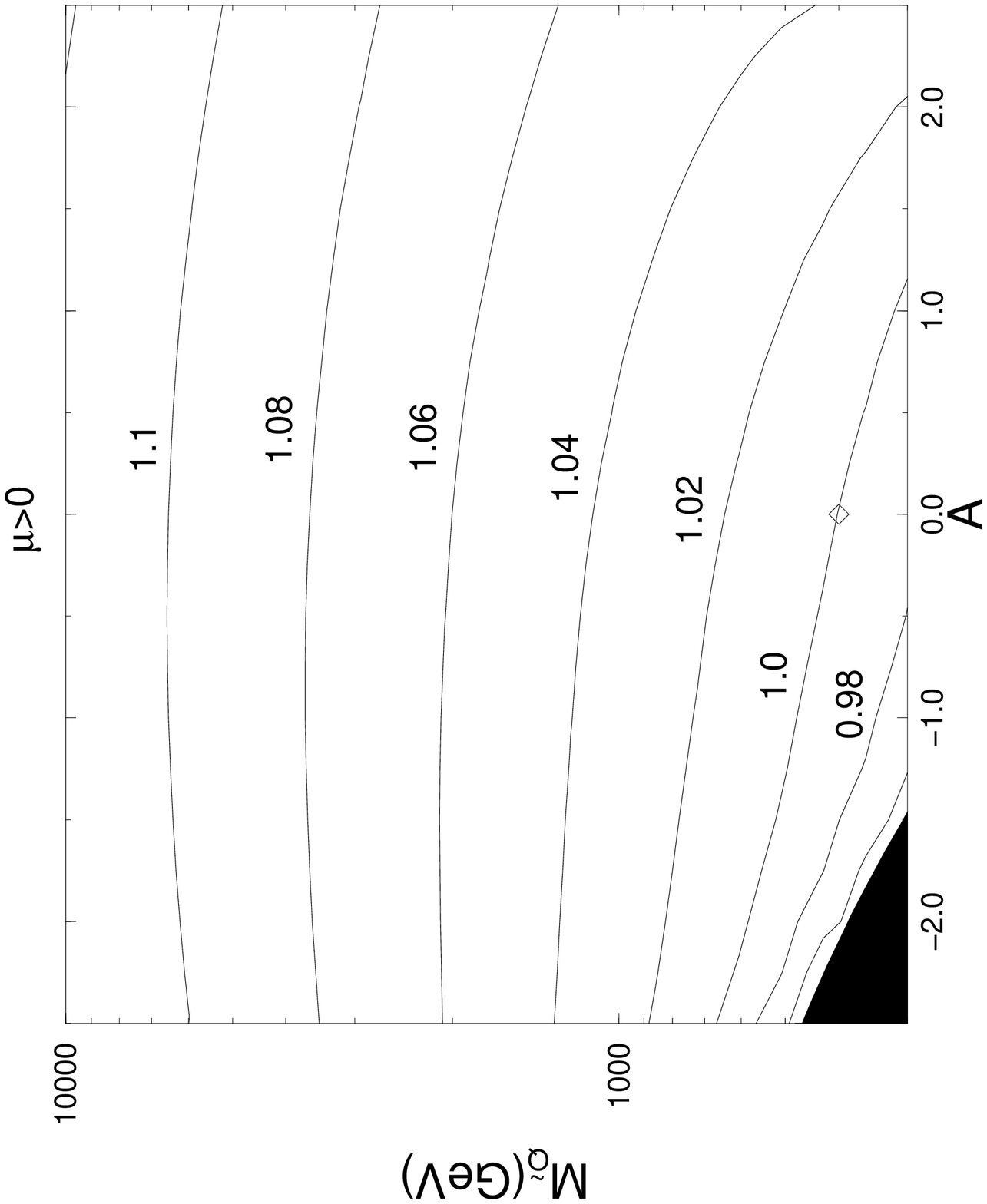}
\end{figure}

\newpage

\begin{figure}[htb]
\includegraphics{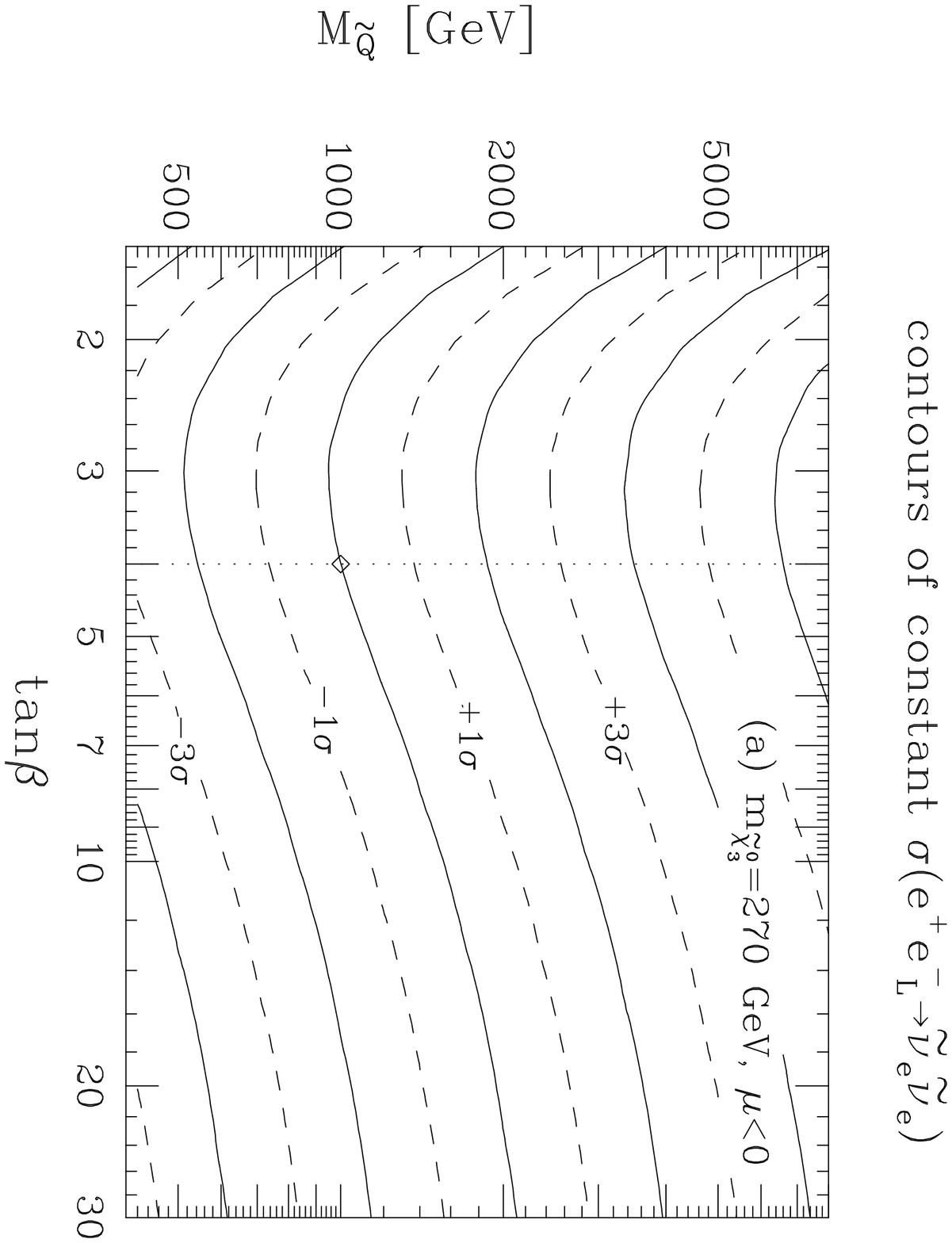}
\includegraphics{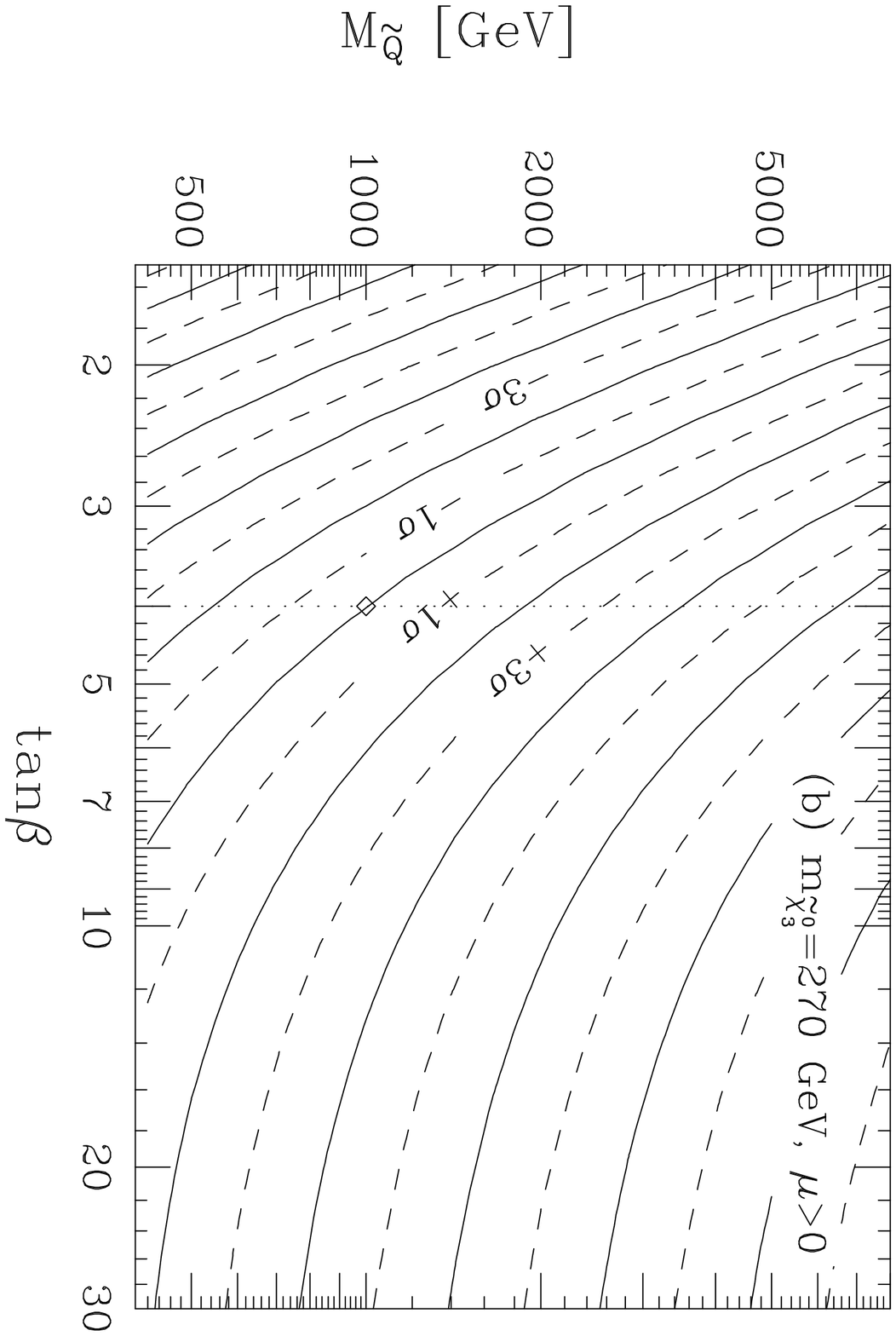}
\includegraphics{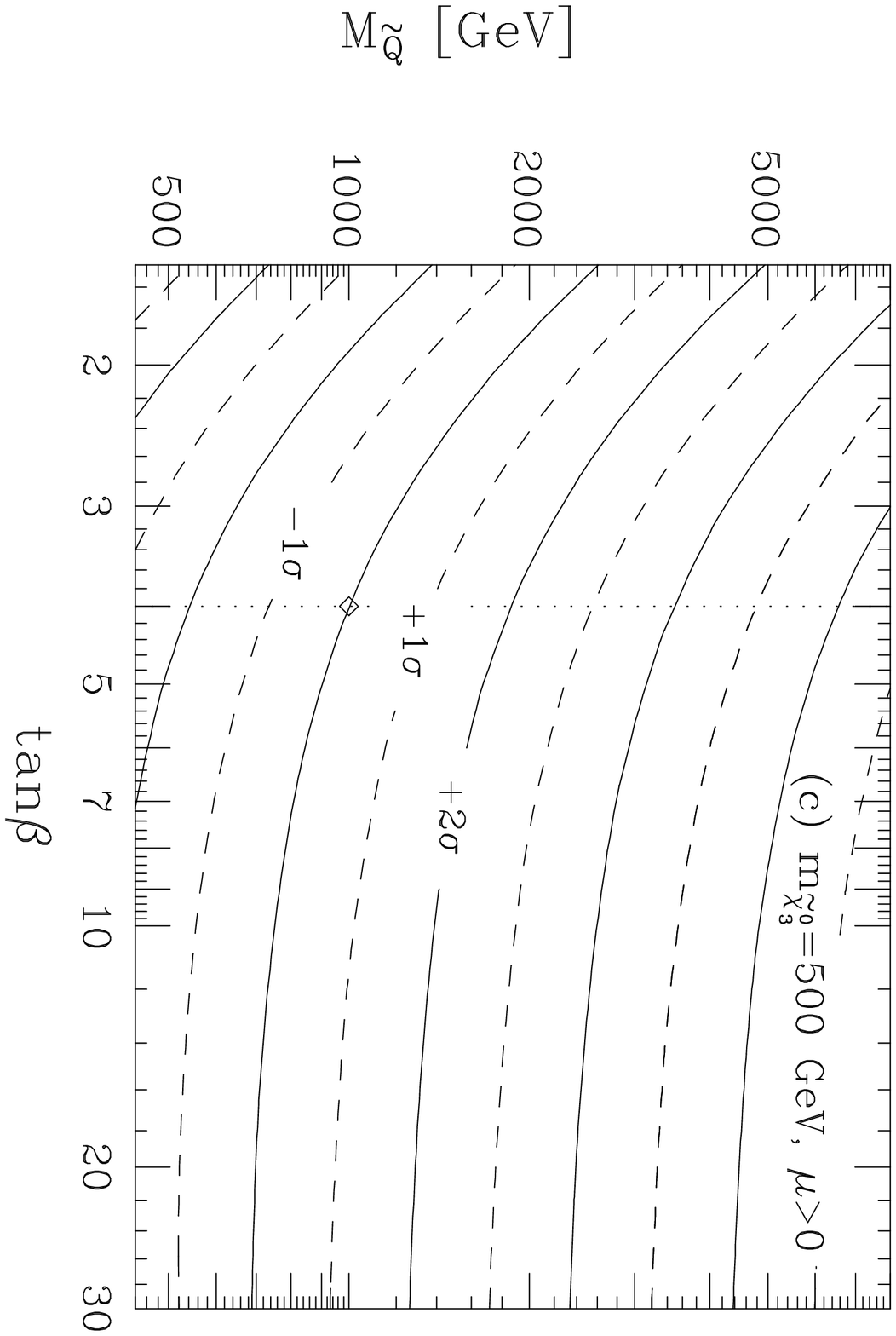}
\caption{The constraint on $M_{\tilde{Q}}$ and $\tan\beta$ coming from
$\sigma(e^-e^+\rightarrow\tilde{\nu}_e\tilde{\nu}_e^*\rightarrow
e^-e^+\tilde{\chi}^+_1\tilde{\chi}^-_1)$, with $\int dt L=100~{\rm
fb}^{-1}$. The central value is taken as $M_{\tilde{Q}}$=1000 GeV and
$\tan\beta(M_Z) =4$. Plotted are the one-loop corrected constant
sneutrino production cross-section corresponding to a given number of
$\sigma$ fluctuations from the central value.  The masses are fixed
and the mass uncertainties are not taken into account.  For (a)
$m_{\tilde\chi^0_3}=270$ GeV, $\mu<0$, (b) $m_{\tilde\chi_3^0}=270$
GeV, $\mu>0$ and (c) $m_{\tilde\chi_3^0}=500$ GeV, $\mu>0$.}
\end{figure}

\newpage

\begin{figure}[htb]
\includegraphics{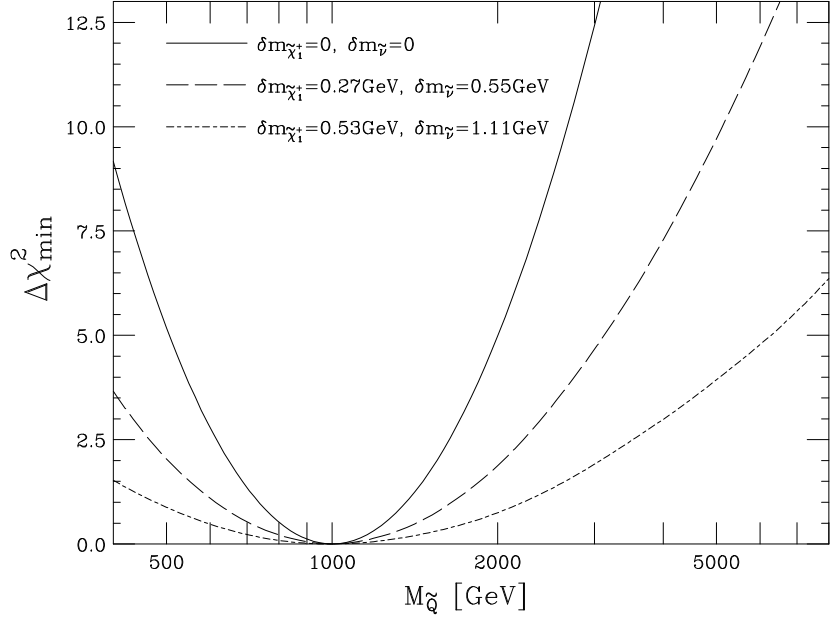}
\caption{$\Delta\chi^2_{\rm min}$ vs. $M_{\tilde Q}$ with $\mu<0$ and
fixed $\tan\beta$, $m_{\tilde{\chi}^0_1}$ and $m_{\tilde{\chi}^0_3}$,
but allowing $m_{\tilde{\chi}^+_1}$ and $m_{\tilde{\nu}_e}$ to vary
freely. The assumed mass error is ($\delta m_{\tilde{\chi}^+_1}$,
$\delta m_{\tilde\nu}$) = (0,0) for the solid line, (0.27~GeV,
0.55~GeV) for the dashed line, and (0.53~GeV, 1.11~GeV) for the short
dashed line.  The mass errors for the short dashed line are obtained
by statistically scaling up the results of Ref.~\protect\cite{BMT}
from 20 fb$^{-1}$ to 100 fb$^{-1}$.}
\end{figure}

\end{document}